\documentclass[aps,twocolumn,prd,showpacs,nofootinbib]{revtex4}

\usepackage{bm}
\usepackage{latexsym}
\usepackage{amsfonts,amssymb,amsmath}
\usepackage{graphicx,epsfig}
\usepackage{psfrag}
\usepackage{subfigure}
\usepackage{ulem}
\usepackage{hyperref}
\usepackage{color}
\usepackage{appendix}

\newcommand{\ie}{{i.e.~}}

\newcommand{\dd}{\mathrm{d}}

\newcommand{\sss}[1]{{\scriptscriptstyle{#1}}}

\newcommand{\uPl}{\mathrm{Pl}}

\newcommand{\uS}{\mathrm{S}}

\newcommand{\usssS}{\sss{\uS}}

\newcommand{\usssPl}{\sss{\uPl}}

\newcommand{\nS}{n_\usssS}

\newcommand{\setR}{\mathbb{R}}

\newcommand{\Mp}{M_\usssPl}

\newcommand{\efolds}{$e$-folds~}

\newcommand{\beq}{\begin{equation}}
\newcommand{\eeq}{\end{equation}}
\newcommand{\bea}{\begin{eqnarray}}
\newcommand{\eea}{\end{eqnarray}}

\newlength{\wsingfig}
\newlength{\wdblefig}
\newlength{\wquadfig}
\newlength{\wtriplefig}
\newcommand{\Eq}[1]{Eq.~(\ref{#1})}
\newcommand{\Eqs}[1]{Eqs.~(\ref{#1})}
\newcommand{\Fig}[1]{Fig.~{\ref{#1}}}

\newcommand{\Ref}[1]{Ref.~{\cite{#1}}}
\newcommand{\Refs}[1]{Refs.~{\cite{#1}}}
\newcommand{\Sec}[1]{Sec.~\ref{#1}}

\newcommand{\Subsec}[1]{sub-Sec.~\ref{#1}}

\newcommand{\App}[1]{Appendix.~\ref{#1}}

\newcommand{\mapPlanck}{\includegraphics[width=0.7cm,height=0.3cm,trim=-1cm 2cm 0
0]{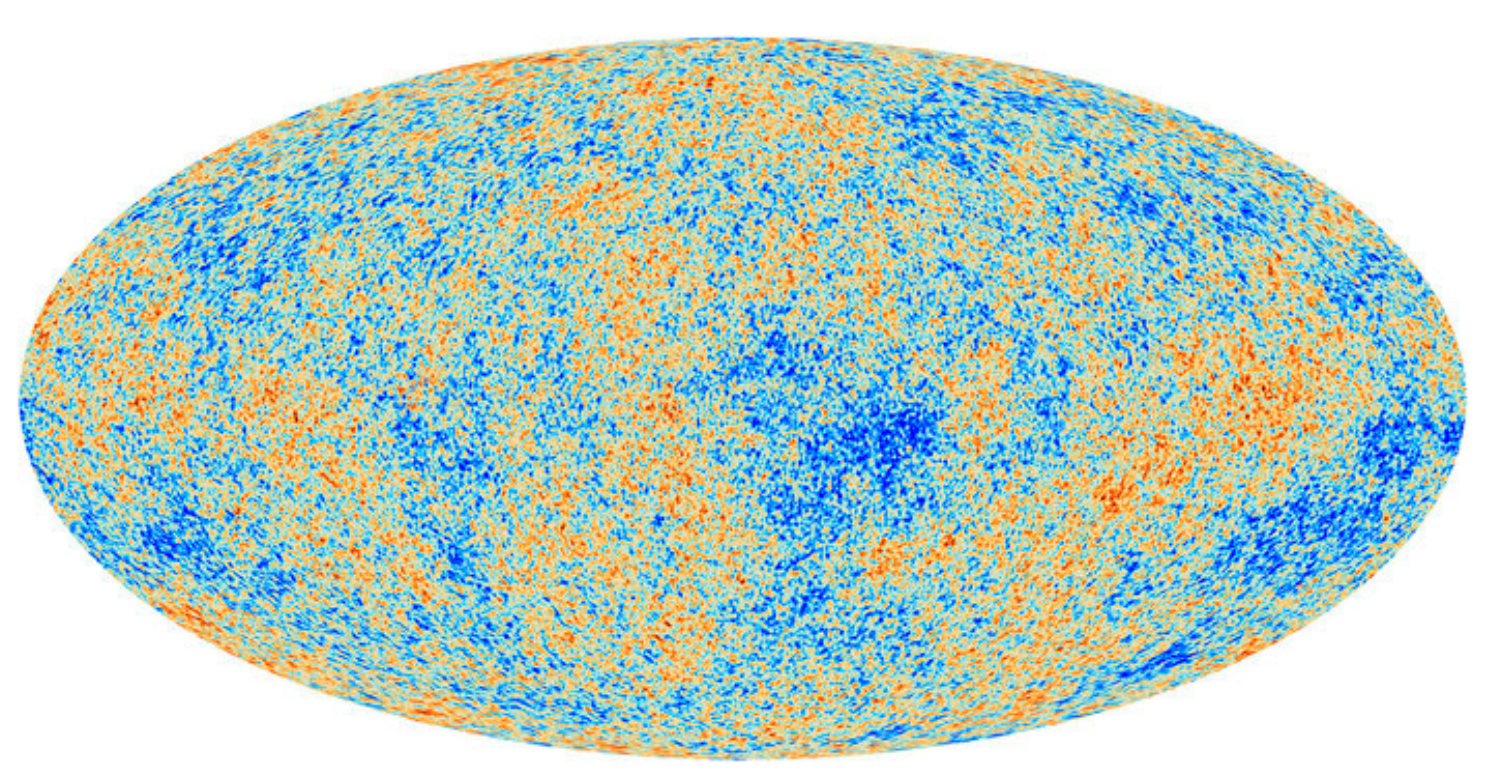}}
\newcommand{\ketmapPlanck}{\vert\mapPlanck\rangle}

\newcommand{\map}{\includegraphics[width=0.7cm,height=0.3cm,trim=-1cm 2cm 0
0]{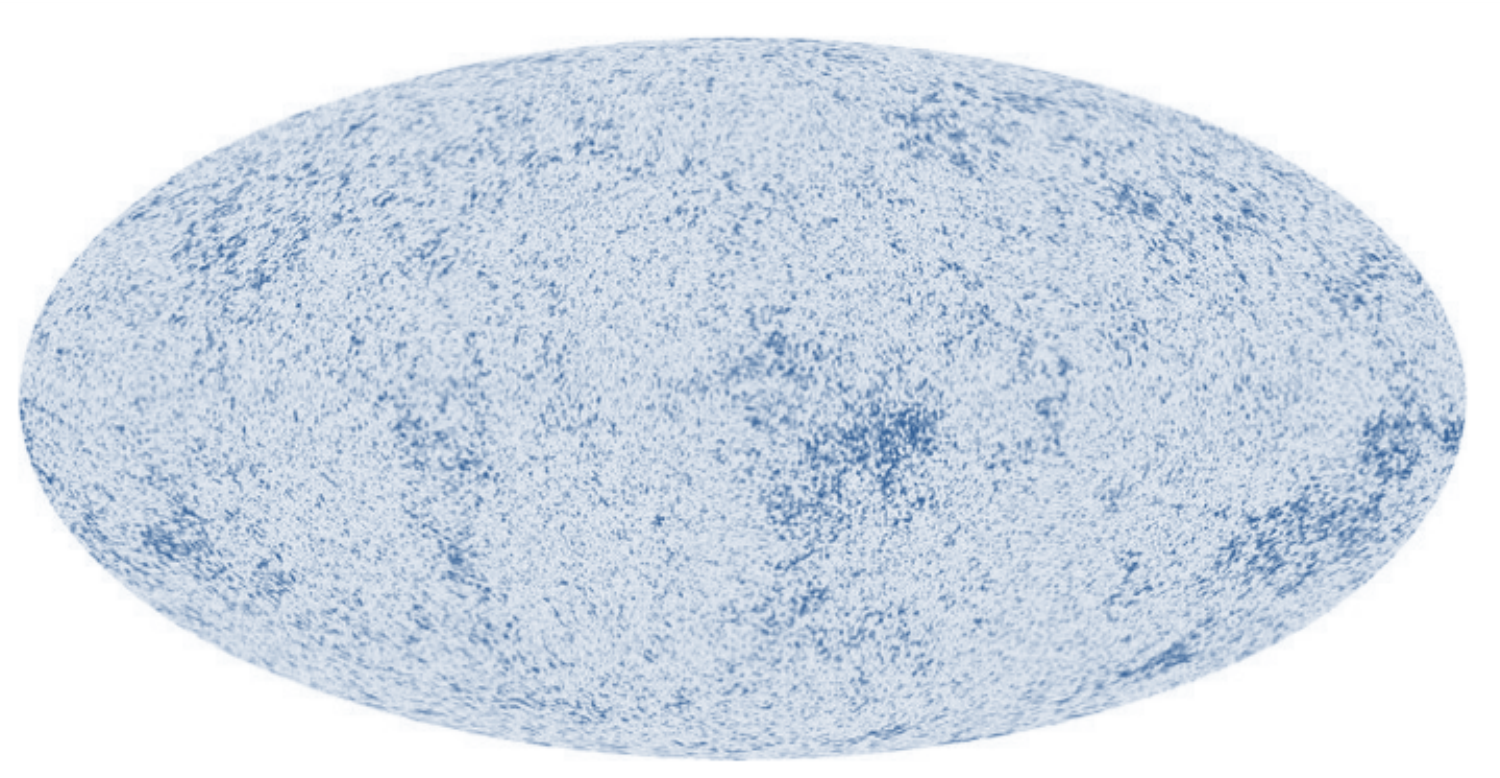}}
\newcommand{\ketmap}{\vert\map\rangle}

\hypersetup{
    bookmarks=true,         
    unicode=false,          
    pdftoolbar=true,        
    pdfmenubar=true,        
    pdffitwindow=false,     
    pdfstartview={FitH},    
    pdftitle={My title},    
    pdfauthor={Author},     
    pdfsubject={Subject},   
    pdfcreator={Creator},   
    pdfproducer={Producer}, 
    pdfkeywords={keyword1} {key2} {key3}, 
    pdfnewwindow=true,      
    colorlinks=true,       
    linkcolor=red,          
    citecolor=cyan,        
    filecolor=magenta,      
    urlcolor=green,           
    linktocpage=true
}

\begin{document}

\title{Obstructions to Bell CMB Experiments}

\author{J\'er\^ome Martin} \email{jmartin@iap.fr}
\affiliation{Institut d'Astrophysique de Paris, UMR 7095-CNRS,
Universit\'e Pierre et Marie Curie, 98 bis boulevard Arago, 75014
Paris, France}

\author{Vincent Vennin} \email{vennin@apc.univ-paris7.fr}
\affiliation{Laboratoire Astroparticule et Cosmologie, Universit\'e Denis Diderot Paris 7, 75013 Paris,
France}
\affiliation{Institute of Cosmology \& Gravitation, University of Portsmouth, Dennis Sciama Building, Burnaby Road, Portsmouth, PO1 3FX, United Kingdom}

\date{\today}

\begin{abstract}
  We present a general and systematic study of how a Bell experiment on the cosmic microwave background could be carried out. We introduce different classes of pseudo-spin operators and show that, if the system is placed in a two-mode squeezed state as inflation predicts, they all lead to a violation of the Bell inequality. However, we also discuss the obstacles that one faces in order to realize this program in practice and show that they are probably insurmountable. We suggest alternative methods that could reveal the quantum origin of cosmological structures without relying on Bell experiments.
\end{abstract}

\pacs{98.80.Cq, 98.80.Qc, 03.67.-a, 03.67.Mn}
\maketitle

\section{Introduction}
\label{sec:intro}

According to inflation~\cite{Starobinsky:1980te, Sato:1980yn,
  Guth:1980zm, Linde:1981mu, Albrecht:1982wi, Linde:1983gd}, galaxies,
clusters of galaxies as well as the cosmic microwave background (CMB)
anisotropies~\cite{Ade:2013zuv,Planck:2013jfk,Ade:2013ydc,Ade:2015xua,Ade:2015lrj,Ade:2015ava}
are of quantum mechanical origin. Conceptually, this fundamental
insight is revolutionary since it means that the structures in
our Universe are, ultimately, nothing but quantum fluctuations
stretched over cosmological
distances~\cite{Mukhanov:1981xt,Mukhanov:1982nu,Starobinsky:1982ee,Guth:1982ec,Hawking:1982cz,Bardeen:1983qw}
(for reviews, see
\Refs{Mukhanov:1990me,Martin:2003bt,Martin:2004um,Martin:2007bw,Linde:2007fr,Sriramkumar:2009kg,Vennin:2015eaa}). Moreover,
the quantum state of these fluctuations is a two-mode squeezed
state~\cite{Grishchuk:1990bj,Grishchuk:1992tw} which is an entangled
state, namely a quantum state that possesses highly non-classical
properties. This opens up the possibility to observe genuine quantum
effects from the very early Universe in the sky, a fascinating prospect indeed.

However, it has been suggested, maybe surprisingly given the non-classical
character of the CMB quantum state, that achieving this goal may be difficult. 
The reason is that the quantum-mechanical phase space of cosmological perturbations is made of two non-commuting conjugated variables, the so-called growing and decaying modes. During inflation,
the decaying mode decays exponentially. This is why it is, a priori,
impossible to measure any quantity related to its amplitude, hence to its commutator with the growing mode. Detecting quantum correlations via this commutator therefore seems intractable in practice. For instance, its has been shown~\cite{Polarski:1995jg,Martin:2015qta} that the two-point quantum correlation functions involving the growing mode only, \ie the ones that are realistically observable, are, in practice, indistinguishable from their ``classical'' counterparts.

Nevertheless, the question of designing Bell CMB experiments~\cite{Campo:2005sv,
  Maldacena:2015bha, Martin:2016tbd, Choudhury:2016cso,Chen:2017cgw,Kanno:2017dci} is worth investigating for the three following reasons. First, the qualitative argument mentioned above is rather vague and it is important to study how it manifests itself in a concrete and explicit attempt to carry out a Bell experiment on the CMB. Only then can one assess how severe it is and whether it can be circumvented or not. Second, the argument holds for measurements of correlation functions at a single wavenumber $\bm{k}$, while Bell CMB experiments typically involve correlators between quantities calculated at modes $\bm{k}$ and $-\bm{k}$, which makes the argument incomplete. We will see that indeed, Bell inequalities can be violated even in the limit where the decaying mode vanishes. Third, there are non-minimal cases where the decaying mode may be accessible, either because it does not vanish on large scales during inflation if sourced by isocurvature perturbations~\cite{Leach:2001zf} for instance, or if, as proposed in \Ref{Maldacena:2015bha}, it couples to other fields and becomes observable through them. In such cases, it remains to determine how Bell inequality violations
could be extracted from the CMB data.

This is why in this article, we carry out a systematic
study and discussion of all the obstacles one faces when trying to
implement a Bell CMB experiment. This leads us to considerations, for instance, the use of CMB pseudo-spin operators, or the connection between the positivity of the Wigner function and the so-called proper variables, that shed new light on these issues. Based on the results obtained here, we also discuss alternative methods that may reveal the quantum origin of cosmological perturbations without relying on Bell experiments.

The paper is organized as follows. In the next section,
\Sec{sec:motivation}, we briefly review the theory of
inflationary cosmological perturbations of quantum-mechanical origin
and discuss the motivations of the present study. Then, in
\Sec{sec:cmbbell}, we explain how a cosmic Bell experiment could
be designed. In a first step, see \Subsec{subsec:bellspin}, we
briefly review how the Bell experiment is usually carried out in
a standard context. Then, we proceed by analogy. In
\Subsec{subsec:bellfakespin}, we explain how a dichotomic
variable can be extracted out from a continuous variable system,
introduce a first class of pseudo-spin operators and show that they
lead to a violation of the Bell inequality. In
\Subsec{subsec:alternative} and
\Subsec{subsec:measurablespin}, we consider two other sets of
possible pseudo-spin operators also leading to a Bell inequality
violation. In \Sec{sec:discussion}, we discuss and interpret
these results. In \Subsec{eq:bivwigner}, we demonstrate that our
results are consistent with a known theorem stating that, if the
Wigner function of the system is positive (which is the case in
cosmology), then a Bell inequality violation can occur only for
improper dynamical variables.  In
\Subsec{subsec:spinmeasurability}, we study whether the pseudo-spin variables can be measured experimentally and conclude that the
answer is probably negative. In \Subsec{sec:decoherence}, we study the robustness of our results to decoherence effects.
Finally, in \Sec{sec:conclusions},
we present our conclusions. The last section of the paper is
\App{sec:correlSx} where we detail the calculation of the
correlation function of one of the $\hat{\cal{S}}_x$ operator introduced in
this article.

\section{Motivation}
\label{sec:motivation}

Inflation produces two types of fluctuations, scalars and tensors. In
the following, we restrict ourselves to scalar perturbations since they are the only ones observed so far and they decouple from the tensor ones. However, this
does not limit in any way the generality of our argument since tensor
modes could be studied in a similar fashion. The evolution of scalar
perturbations is controlled by the following
Hamiltonian~\cite{Mukhanov:1981xt,Mukhanov:1990me}
\begin{align}
\label{eq:hami}
\hat{H} &= \int_{\mathbb{R}^{3}} \dd ^3{\bm k}
\biggl[\frac{k}{2}\left(\hat{c}_{\bm k}\hat{c}_{\bm k}^{\dagger}
+\hat{c}_{-{\bm k}}\hat{c}_{-{\bm k}}^{\dagger}
\right)
\nonumber \\ & 
-\frac{i}{2}\frac{z'}{z}
\left(\hat{c}_{\bm k}\hat{c}_{-{\bm k}}
-\hat{c}_{-{\bm k}}^{\dagger}\hat{c}_{\bm k}^{\dagger}\right)
\biggr],
\end{align}
where $\hat{c}_{\bm k}$ and $\hat{c}_{\bm k}^{\dagger}$ are
respectively the creation and annihilation operators satisfying the
standard commutation relation
$[\hat{c}_{\bm k},\hat{c}_{\bm p}^{\dagger}]=\delta ({\bm k}-{\bm
  p})$.
In the above expression, the quantity $z$ is defined to be
$z\equiv a\Mp\sqrt{2\epsilon_1}$, where $a(t)$ is the
Friedmann-Lemaitre-Robertson-Walker (FLRW) scale factor, $\Mp$ the
reduced Planck mass and $\epsilon_1$ the first Hubble flow parameter,
$\epsilon_1\equiv -\dot{H}/H^2$, $H=\dot{a}/a$ being the Hubble
parameter. A dot denotes a derivative with respect to cosmic time $t$
while a prime represents a derivative with respect to conformal time
$\eta$ with ${\rm d}t=a{\rm d}\eta$. The creation and annihilation
operators are related to the scalar curvature perturbations
$\hat{\zeta}_{\bm k}$ through
$\hat{v}_{\bm k}=(\hat{c}_{\bm k}+\hat{c}_{-{\bm
    k}}^{\dagger})/\sqrt{2k}$
and
$\hat{p}_{\bm k}=-i\sqrt{k/2}(\hat{c}_{\bm k}-\hat{c}_{-{\bm
    k}}^{\dagger})$
by $\hat{\zeta}_{\bm k}=\hat{v}_{\bm k}/z$ and
$\hat{\zeta}'_{\bm k}=\hat{p}_{\bm k}/z$. The quantity
$\hat{v}_{\bm k}$ is the so-called Mukhanov-Sasaki variable. We also
introduce the quantities $\hat{q}_{\bm k}$ and $\hat{\pi}_{\bm k}$
given by
$\hat{q}_{\bm k}=(\hat{c}_{\bm k}+\hat{c}_{{\bm
    k}}^{\dagger})/\sqrt{2k}$
and
$\hat{\pi}_{\bm k}=-i\sqrt{k/2}(\hat{c}_{\bm k}-\hat{c}_{{\bm
    k}}^{\dagger})$.
The advantage of these variable is that they are defined for a fixed
wavenumber $\bm{k}$ (while $\zeta_{\bm k}$ mixes ${\bm k}$ and $-{\bm k}$) and they are hermitian. In
this sense, they really play the role of the position and momentum at
the scale ${\bm k}$. The relation between these two sets of variables
is easily obtained, see for instance Eqs.~(51) and (52) of
\Ref{Martin:2015qta}, and is given by
\begin{align}
\label{eq:linkvq}
\hat{v}_{\bm k}&=\frac12\left[\hat{q}_{\bm k}+\hat{q}_{-{\bm k}}
+\frac{i}{k}\left(\hat{\pi}_{\bm k}-\hat{\pi}_{-{\bm k}}\right)\right],\\
\label{eq:linkpq}
\hat{p}_{\bm k}&=\frac{1}{2i}\left[k\left(\hat{q}_{\bm k}-\hat{q}_{-{\bm k}}\right)
+i\left(\hat{\pi}_{\bm k}+\hat{\pi}_{-{\bm k}}\right)\right].
\end{align}
The Hamiltonian given by \Eq{eq:hami} is well known and
represents a collection of parametric oscillators. The above
discussion shows that the full Hilbert space of the system ${\cal E}$
can be factorized into independent products of Hilbert spaces for
modes ${\bm k}$ and $-{\bm k}$,
\bea
\label{eq:HilbertSpace}
{\cal E}=\Pi_{k\in\mathbb{R}^{3+}}{\cal E}_{\bm k}\otimes {\cal
  E}_{-{\bm k}}\, .
 \eea
In other words, the CMB fluctuations can be viewed as a product of independent bipartite
systems. If, initially, the quantum state is the vacuum state
$\vert 0_{\bm k},0_{-{\bm k}}\rangle$, then, under the
Hamiltonian~(\ref{eq:hami}), it evolves into a two-mode squeezed
state~\cite{Grishchuk:1990bj,Grishchuk:1992tw,Polarski:1995jg,Martin:2004um,Martin:2007bw}
\begin{align}
\vert \Psi_{2\, {\rm sq}}\rangle & = \frac{1}{\cosh r_k} \sum _{n=0}^{\infty}
{\rm e}^{-2in\varphi_k}\tanh ^n r_k \vert n_{\bm k},n_{-{\bm k}}\rangle, 
\label{eq:qstate}
\end{align}
where $\vert n_{\bm k}\rangle $ is an eigenvector of the particle
number operator,
$N_{\bm k}=\hat{c}_{{\bm k}}^{\dagger}\hat{c}_{{\bm k}}$. In the above
expression, $r_k$ and $\varphi_k$ are respectively the squeezing
parameter and squeezing angle. They are functions of time that depend on the details of the inflationary background, but are such that on sub-Hubble scales, \ie when $k\gg aH$, $r_k\rightarrow 0$ and $\varphi_k \rightarrow -3\pi/4$, while on super-Hubble scales, \ie when $k \ll aH$, $r_k \rightarrow \infty$ and $\varphi_k \rightarrow -\pi$.\footnote{\label{footnote:varphi:convention}In \Ref{Martin:2015qta},
  the two-mode squeezed state was written \begin{align} \vert
    \Psi_{2\, {\rm sq}}\rangle & = \frac{1}{\cosh r_k} \sum
    _{n=0}^{\infty} {\rm e}^{2in\varphi_k}(-1)^n \tanh ^n r_k \vert
    n_{\bm k},n_{-{\bm k}}\rangle,
\label{eq:qstatediscord}
\end{align}
which implies a different definition of the squeezing angle, namely
$\varphi_k \rightarrow -(\varphi_k+\pi/2)$. With the convention of
\Ref{Martin:2015qta}, the squeezing angles goes to $\pi/2$ on large
scales. Moreover in \Ref{Martin:2016tbd}, the symmetries of the problem allow the analysis to be restricted to the range $\varphi_k \in[0,\pi/4]$. During inflation, $\varphi_k\in[-\pi,-3\pi/4]$ and if one wants to use the results of \Ref{Martin:2016tbd}, this can be done by changing $\varphi_k \rightarrow \varphi_k+\pi$.} More precisely, on super-Hubble scales, $r_k$ is of order the number of \efolds $N\equiv \ln a$ spent outside the Hubble radius, which is $\sim 50$ for the modes probed in the CMB.
 
The classical limit in quantum mechanics is subtle and this question
is perfectly illustrated in the case of a two-mode squeezed
state. Usually, this state is considered as the prototype of a
``non-classical'' quantum state. In this language, a ``classical''
state would typically be a coherent state, namely a state whose Wigner
function follows the trajectory of the classical system in phase space
with minimum dispersion. On the contrary, the Wigner function of a
squeezed state has a large extension in phase space (to be more
precise in a specific direction). Moreover, the extension is all the
greater when the squeezing parameter is large. And this is precisely
the case in cosmology where the values of $r_k$ are much larger than
what can be achieved in the laboratory with conventional systems. In
this sense, the sky is placed in a highly ``quantum-mechanical''
state.

On the other hand, a system is also said to be classical if its Wigner
function is positive. Indeed, it is well known that interference
terms, which are typical signatures of a quantum behavior, cause
oscillations in the Wigner function and make it negative for some
values of position and momentum. But a squeezed state is a Gaussian
state and, therefore, its Wigner function is always positive. In this
sense, and contrary to the previous criterion, the system is classical.

However, as is apparent in \Eq{eq:qstate}, the two-mode squeezed
state is also an entangled state. Clearly, entanglement is a genuine
quantum-mechanical feature and, therefore, this feature brings us
back to the first conclusion. Moreover, the larger $r_k$, the larger
the entanglement as confirmed by a calculation of the quantum
discord\footnote{For pure states such as \Eq{eq:qstate}, the quantum discord becomes a measure of quantum entanglement. More specifically in that case, it equals the entropy of entanglement.}~\cite{Henderson:2001,Zurek:2001} (for a review see \Ref{Adesso:2016ygq}) performed in
\Ref{Martin:2015qta}
\begin{align}
\label{eq:discord}
\delta \left({\bm k},-{\bm k}\right)
&=\cosh ^2r_k\log_2\left(\cosh ^2 r_k\right)
\nonumber \\ &
-\sinh ^2r_k\log_2\left(\sinh ^2 r_k\right)
\nonumber \\ 
&\simeq 
\frac{2}{\ln 2} r_k-2+\frac{1}{\ln 2}
+{\cal O}\left({\rm e}^{-2r_k}\right)\, ,
\end{align}
where the last expansion is valid for large values of the squeezing
parameter. 

\begin{figure*}[t]
\begin{center}
\includegraphics[width=13cm]{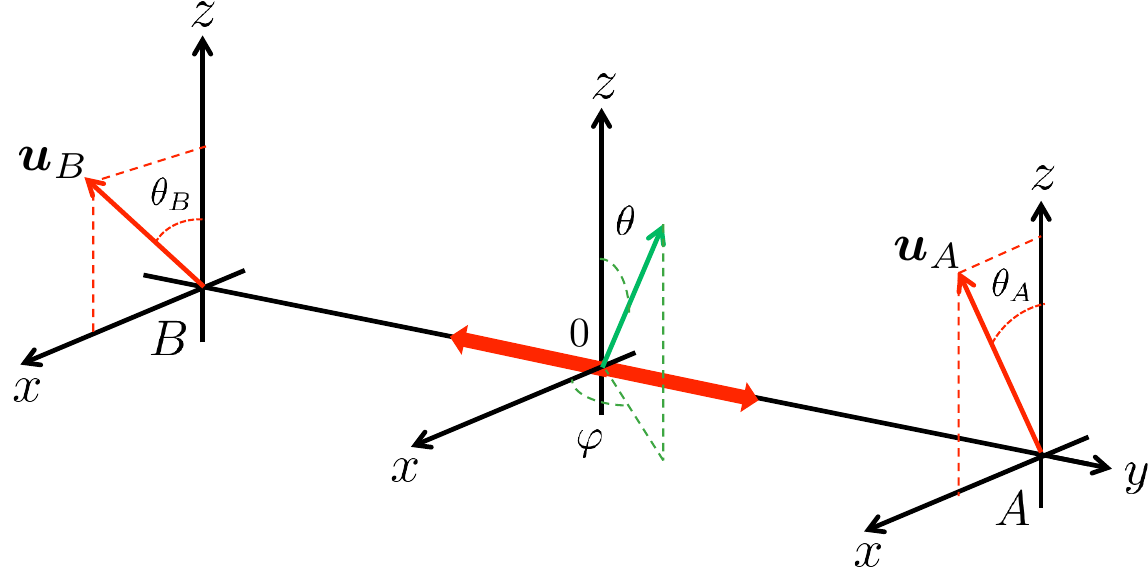}
\end{center}
\caption{Standard setup for Bell experiment. Two spin particles are emitted
  at the origin $0$ and travel in opposite directions. Their spin is then
  measured at $A$ and $B$ along the directions ${\bm u}_A$ and ${\bm
    u}_B$.}
\label{fig:bellstandard}
\end{figure*}

If the CMB does not have a classical behavior, this means that, at
least in principle, one should be able to design experiments allowing
us to exhibit its non-classical properties. One usually characterizes
the CMB by the correlation functions of curvature
perturbations. If the system is really ``non classical'', then these
correlation functions should differ according to whether one
calculates them with a two-mode squeezed state or with a classical
state. Here, by classical state, we mean a state which contains
classical correlations only and has zero discord~\cite{Werner:1989zz}. In \Ref{Martin:2015qta}, it has been shown that such a state can be chosen so as to
exactly reproduce the two-point correlation function
$\langle \hat{\zeta}(\eta, {\bm x})\hat{\zeta}(\eta, {\bm y})\rangle$.
But, in accordance with the fact that the discord of the system is
large, this would then imply correlation functions
$\langle \hat{\zeta}(\eta, {\bm x})\hat{\zeta}'(\eta, {\bm
  y})+\hat{\zeta}'(\eta, {\bm x})\hat{\zeta}(\eta, {\bm y})\rangle$
and/or
$\langle \hat{\zeta}'(\eta, {\bm x})\hat{\zeta}'(\eta, {\bm
  y})\rangle$
that strongly differ from their two-mode squeezed state
counterparts. At this stage, one would therefore be tempted to claim
that, indeed, quantum signatures can be detected in the sky. However, in practice, the quantity
$\hat{\zeta}'(\eta, {\bm x})$ is not observable since this is the
decaying mode. Importantly at the conceptual level, we see that the
argument is not a theoretical argument but a practical one. In principle, the difference between two-point correlation functions calculated in a
classical or quantum-mechanical contexts exist and are observable but,
in practice, due to the smallness of the decaying mode amplitude, this
is hidden to us, probably for ever.

If one considers higher correlation functions however (such as the four-point correlation function and higher), even when relying on the growing mode only, there is no classical state that can reproduce the predictions of the two-mode squeezed state~\cite{Martin:2015qta}. This is consistent with the theorem~\cite{PhysRevLett.105.030501, PhysRevA.87.012119, PhysRevA.90.022328} stating that the only classical Gaussian states are product states, \ie uncorrelated states. Since the CMB statistics is constrained to be Gaussian or almost Gaussian~\cite{Ade:2015ava}, and since the bipartite systems of \Eq{eq:HilbertSpace} are necessarily correlated [if not one can simply perform a phase space transformation similar to Eqs.~(47) and~(48) of \Ref{Martin:2015qta} to obtain a correlated system], these correlations must therefore possess non-vanishing discord. For a pure state, non-vanishing discord is equivalent to entanglement and this implies that Bell inequalities can (at least in principle) be violated. This is the possibility we investigate in this work. Otherwise, this implies that the CMB is placed in a mixed, decohered state, another interesting possibility that we will discuss in a separate article, see also \Subsec{sec:decoherence}.
\section{Bell CMB experiment with pseudo-spin operators}
\label{sec:cmbbell}
\subsection{Bell inequality with spins}
\label{subsec:bellspin}
In order to see how we could design a Bell CMB experiment, let us
first, very briefly, recall how this is done in a conventional
situation, the so-called Clauser, Horne, Shimony and Holt (CHSH)
setup~\cite{Clauser:1969ny}. The idea is to consider a bipartite
system whose Hilbert space is written as
${\cal H}={\cal H}_A\otimes {\cal H}_B$.  Typically, $A$ and $B$ are
two particles whose spin along the $z$ direction are correlated, see
Fig.~\ref{fig:bellstandard}. The state of the system is assumed to be
\begin{align}
\label{eq:qstatespin}
\vert \Psi \rangle =\frac{1}{\sqrt{2}}
\left(\vert +,-\rangle -\vert -,+\rangle \right),
\end{align}
where $\vert \pm\rangle $ are eigenstates of $\hat{S}_z$ with,
respectively, eigenvalues $\pm 1$. The spin of system $A$ is measured
along the direction characterized by the angle $\theta_A$ in the
$(x,z)$ plane, namely along the vector ${\bm u}_A$ and the spin of
system $B$ is measured along the direction ${\bm u}_B$, see
Fig.~\ref{fig:bellstandard}. In general, if ${\bm S}\cdot {\bm u}$ is
the spin operator along the direction
${\bm u}=(\sin \theta \cos \varphi, \sin \theta \sin \varphi, \cos
\theta)$,
the eigenstates of ${\bm S}\cdot {\bm u}$ are
$\vert +_{\bm u}\rangle=\cos(\theta/2)e^{-i\varphi/2}\vert +\rangle
+\sin(\theta/2)e^{i\varphi/2}\vert-\rangle$
and
$\vert -_{\bm u}\rangle =-\sin(\theta/2)e^{-i\varphi/2}\vert +\rangle
+\cos(\theta/2)e^{i\varphi/2}\vert-\rangle$
with eigenvalues $\pm 1$. Then, one can introduce the Bell
operator
\begin{align}
\label{eq:standardB}
\hat{{\cal B}}_{_{\rm CHSH}}(A,B)
&={\bm u}_A\cdot \hat{{\bm S}}_A
\otimes {\bm u}_B\cdot \hat{{\bm S}}_B
+{\bm u}_A\cdot \hat{{\bm S}}_A
\otimes {\bm u}'_B\cdot \hat{{\bm S}}_B
\nonumber \\ &
+{\bm u}'_A\cdot \hat{{\bm S}}_A
\otimes {\bm u}_B\cdot \hat{{\bm S}}_B
-{\bm u}'_A\cdot \hat{{\bm S}}_A
\otimes {\bm u}'_B\cdot \hat{{\bm S}}_B,
\end{align}
where ${\bm u}_A$, ${\bm u}'_A$, ${\bm u}_B$ and ${\bm u}'_B$ are four
different vectors, all located in the $(x,z)$ plane (and, therefore,
with vanishing azimuthal angles). Then, one has to calculate the mean
value of the Bell operator in the
state~(\ref{eq:qstatespin}). One can show that
$\langle \hat{{\cal B}}_{_{\rm
    CHSH}}(A,B)\rangle=E(\theta_A,\theta_B)+E(\theta_A,\theta_B')+E(\theta_A',\theta_B)-E(\theta_A',\theta_B')$
with
$E(\theta_A,\theta_B)\equiv \langle {\bm u}_A\cdot \hat{{\bm S}}_A
\otimes {\bm u}_B\cdot \hat{{\bm S}}_B\rangle
=-\cos(\theta_A-\theta_B)$.
If, for instance, one chooses $\theta_A-\theta_B=\pi/4$,
$\theta_A-\theta_B'=\theta_A'-\theta_B=-\pi/4$ and
$\theta_A'-\theta_B'=-3\pi/4$, then
$\langle \hat{{\cal B}}_{_{\rm CHSH}}(A,B)\rangle =-2\sqrt{2}$. Since
$\vert \langle \hat{{\cal B}}_{_{\rm CHSH}}(A,B)\vert >2$, the Bell
inequality is violated and this
cannot be accounted for in a theory with local realism. As is
well known, this has been experimentally
confirmed~\cite{Aspect:1982fx,Aspect:1981nv,Weihs:1998gy,Hensen:2015ccp}. Very
recently, this has even been observed with a set-up where the
detectors are controlled by the light coming from distant
stars~\cite{Handsteiner:2016ulx}.
 
Our goal is now to design a similar approach but with the CMB.

\subsection{Banaszek-Wodkiewicz (BW) spin operators}
\label{subsec:bellfakespin}

The first difficulty that we meet is that we deal with a continuous
variable system. Indeed $\hat{\zeta} _{\bm k}$ (or the Mukhanov-Sasaki
variable $\hat{v}_{\bm k}$) are continuous complex operators and have
a continuous spectrum, not a discrete one with two eigenvalues
$\pm 1$. However, for any continuous variable system, it is possible
to introduce fictitious or pseudo-spin operators. They have been
discussed by Banaszek and Wodkiewics (BM) in
\Ref{PhysRevLett.82.2009} and Chen, Pan, Hou and Zhang in
\Ref{PhysRevLett.88.040406} and are defined
by~\cite{PhysRevLett.82.2009,PhysRevLett.88.040406,2006FoPh...36..546R,2005PhRvA..71b2103R}
\begin{eqnarray}
\label{eq:defSx1}
\hat{s}_x\left({\bm k}\right)&=& \sum_{n=0}^{\infty}
\left(\vert 2n_{\bm k}+1\rangle \langle 2n_{\bm k}\vert +\vert 2n_{\bm k}\rangle 
\langle 2n_{\bm k}+1\vert \right) \\
\label{eq:defSy1}
\hat{s}_y\left({\bm k}\right) &=& i\sum_{n=0}^{\infty}
\left(\vert 2n_{\bm k}\rangle \langle 2n_{\bm k}+1\vert 
-\vert 2n_{\bm k}+1\rangle 
\langle 2n_{\bm k}\vert \right)\\
\label{eq:defSz1}
\hat{s}_z\left({\bm k}\right) &=& \sum_{n=0}^{\infty}
\left(\vert 2n_{\bm k}+1\rangle \langle 2n_{\bm k}+1\vert 
-\vert 2n_{\bm k}\rangle 
\langle 2n_{\bm k}\vert \right),
\end{eqnarray}
and similar expression for the mode $-{\bm k}$. The states
$\vert n_{\bm k}\rangle$ are the eigenvectors of the particle number
operator already introduced before. It is easy to verify that these
operators satisfy the usual $SU(2)$ commutation relations for a spin, namely
$\left[\hat{s}_x,\hat{s}_y\right]=2i\hat{s}_z$,
$\left[\hat{s}_x,\hat{s}_z\right]=-2i\hat{s}_y$ and
$\left[\hat{s}_y,\hat{s}_z\right]=2i\hat{s}_x$. Moreover, if one
defines a fictitious unit vector
${\bm n}=\left(\sin \theta_n\cos\varphi_n, \sin \theta_n \sin
  \varphi_n, \cos \theta_n\right)$,
then one has $\left({\bm n}\cdot \hat{{\bm s}}\right)^2=\hat{\mathrm{I}}$ which
means that the outcome of a measurement of the Hermitian operator
${\bm n}\cdot {\bm s}$ is, as expected, $\pm 1$. Therefore, we have
achieved a first goal, namely define a dichotomic variable from a
continuous variable system. From this point, one can then proceed by
analogy. We can indeed define the Bell operator by
\begin{align}
\label{eq:defB}
\hat{{\cal B}}_{_{\rm BW}}\left({\bm k},-{\bm k}\right)
&={\bm n}\cdot \hat{{\bm s}}\left({\bm k}\right)
\otimes {\bm m}\cdot \hat{{\bm s}}\left(-{\bm k}\right)
\nonumber \\ & 
+{\bm n}\cdot \hat{{\bm s}}\left({\bm k}\right)
\otimes {\bm m}'\cdot \hat{{\bm s}}\left(-{\bm k}\right)
\nonumber \\ &
+{\bm n}'\cdot \hat{{\bm s}}\left({\bm k}\right)
\otimes {\bm m}\cdot \hat{{\bm s}}\left(-{\bm k}\right)
\nonumber \\ & 
-{\bm n}'\cdot \hat{{\bm s}}\left({\bm k}\right)
\otimes {\bm m}'\cdot \hat{{\bm s}}\left(-{\bm k}\right),
\end{align}
where ${\bm n}$, ${\bm n}'$, ${\bm m}$ and ${\bm m}'$ are four unit
vectors, since this mimics exactly \Eq{eq:standardB}. Then,
one has to calculate the mean value of this operator, not in a state
similar to the one given in \Eq{eq:qstatespin} though but, of
course, in a two-mode squeezed state as \Eq{eq:qstate}. This
gives
\begin{align}
\langle \Psi_{2\, {\rm sq}} \vert \hat{{\cal B}}_{_{\rm BW}}
\left({\bm k},-{\bm k}\right)\vert \Psi_{2\, {\rm sq}}\rangle & =
E\left(\theta_n,\theta_m\right)
+E\left(\theta_n,\theta_{m'}\right)
\nonumber \\ & 
+E\left(\theta_{n'},\theta_m\right)
-E\left(\theta_{n'},\theta_{m'}\right),
\end{align}
where the correlation function $E({\bm n},{\bm m})$ is defined by
\begin{equation}
E({\bm n},{\bm m})=\langle \Psi_{2\,{\rm sq}}\vert 
{\bm n}\cdot \hat{{\bm s}}\left({\bm k}\right)
\otimes {\bm m}\cdot \hat{{\bm s}}\left(-{\bm k}\right)
\vert \Psi_{2\,{\rm sq}}\rangle.
\end{equation}
If we choose all azimuthal angles to be zero (as it is the case in the
standard setup, see the previous sub-section), then one has
${\bm n}\cdot \hat{{\bm s}}=\sin \theta_n \hat{s}_x+\cos
\theta_n\hat{s}_z$ and a straightforward calculation shows that $\langle \Psi_{2\,{\rm sq}}\vert 
\hat{{\bm s}}_x\left({\bm k}\right)
\otimes \hat{{\bm s}}_z\left(-{\bm k}\right)
\vert \Psi_{2\,{\rm sq}}\rangle=0$. As a consequence, one can write
\begin{widetext}
\begin{align}
E\left(\theta_n,\theta_m\right)
& =\langle \Psi_{2\,{\rm sq}}\vert 
\hat{{ s}}_z\left({\bm k}\right)
\otimes \hat{{s}}_z\left(-{\bm k}\right)
\vert \Psi_{2\,{\rm sq}}\rangle\cos \theta_n\cos \theta_m 
+\langle \Psi_{2\,{\rm sq}}\vert 
\hat{{ s}}_x\left({\bm k}\right)
\otimes \hat{{ s}}_x\left(-{\bm k}\right)
\vert \Psi_{2\,{\rm sq}}\rangle
\sin \theta_n\sin \theta_m .
\end{align}
Finally, choosing (for instance) the configuration $\theta_n=0$,
$\theta_{n'}=\pi/2$ and $\theta_{m'}=-\theta_m$ leads to the following
expression
\begin{align}
\langle \Psi_{2\, {\rm sq}} &\vert \hat{{\cal B}}_{_{\rm BW}}
\left({\bm k},-{\bm k}\right)\vert \Psi_{2\, {\rm sq}}\rangle 
 = 2\bigl[\cos\theta_m
\langle \Psi_{2\,{\rm sq}}\vert 
\hat{{ s}}_z\left({\bm k}\right)
\otimes \hat{{ s}}_z\left(-{\bm k}\right)
\vert \Psi_{2\,{\rm sq}}\rangle
+\sin \theta_m
\langle \Psi_{2\,{\rm sq}}\vert 
\hat{{ s}}_x\left({\bm k}\right)
\otimes \hat{{ s}}_x\left(-{\bm k}\right)
\vert \Psi_{2\,{\rm sq}}\rangle\bigr].
\end{align}
One can then optimize the choice of $\theta_m$ in order to obtain the
largest value of
$\langle \Psi_{2\, {\rm sq}} \vert \hat{{\cal B}}_{_{\rm BW}}
\left({\bm k},-{\bm k}\right)\vert \Psi_{2\, {\rm sq}}\rangle$.
This leads to $
\theta_m^{\rm opt}=\arctan
\left[\langle \Psi_{2\,{\rm sq}}\vert 
\hat{{ s}}_x\left({\bm k}\right)
\otimes \hat{{ s}}_x\left(-{\bm k}\right)
\vert \Psi_{2\,{\rm sq}}\rangle / \langle \Psi_{2\,{\rm sq}}\vert 
\hat{{ s}}_z\left({\bm k}\right)
\otimes \hat{{ s}}_z\left(-{\bm k}\right)
\vert \Psi_{2\,{\rm sq}}\rangle\right]
$
and, therefore, for this optimal configuration,
\begin{align}
\label{eq:Bmeanopt}
\langle \Psi_{2\, {\rm sq}} &\vert \hat{{\cal B}}_{_{\rm BW}}
\left({\bm k},-{\bm k}\right)\vert \Psi_{2\, {\rm sq}}\rangle 
=2\sqrt{\langle \Psi_{2\,{\rm sq}}\vert 
\hat{{ s}}_z\left({\bm k}\right)
\otimes \hat{{s}}_z\left(-{\bm k}\right)
\vert \Psi_{2\,{\rm sq}}\rangle^2+\langle \Psi_{2\,{\rm sq}}\vert 
\hat{{ s}}_x\left({\bm k}\right)
\otimes \hat{{ s}}_x\left(-{\bm k}\right)
\vert \Psi_{2\,{\rm sq}}\rangle^2}.
\end{align}
\end{widetext}
For the two-mode squeezed state~(\ref{eq:qstate}), one can show that $\langle \Psi_{2\,{\rm sq}}\vert 
\hat{{ s}}_z\left({\bm k}\right)
\otimes \hat{{ s}}_z\left(-{\bm k}\right)
\vert \Psi_{2\,{\rm sq}}\rangle =1$. Already at this stage, one sees that the Bell inequality is violated
as soon as
$\langle \Psi_{2\, {\rm sq}} \vert \hat{s}_x({\bm k})\otimes 
\hat{s}_x(-{\bm k})
\vert \Psi_{2\, {\rm sq}}\rangle \neq0$. From the state~(\ref{eq:qstate}), one has
\bea
\label{eq:sxsx:BW}
\langle \Psi_{2\,{\rm sq}}\vert 
\hat{{s}}_x\left({\bm k}\right)
\otimes \hat{{ s}}_x\left(-{\bm k}\right)
\vert \Psi_{2\,{\rm sq}}\rangle
= \tanh(2r_k) \cos(2\varphi_k).
\eea
Notice that, in the case of a vanishing squeezing angle, this result
was already derived in \Ref{2005PhRvA..71b2103R} but in a
different context. The case $\varphi_k\neq 0$ is new.
\begin{figure*}[t]
\begin{center}
\includegraphics[width=8cm]{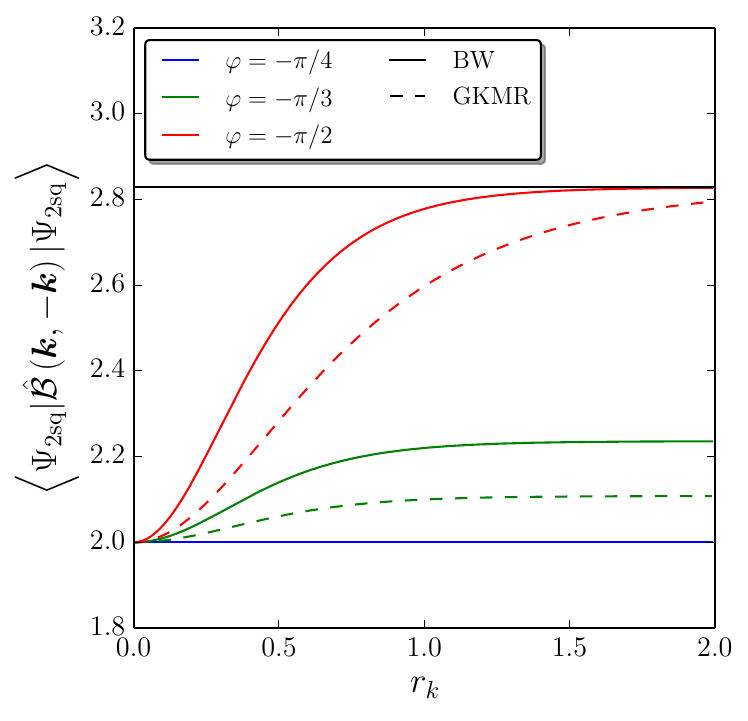}
\includegraphics[width=8.45cm]{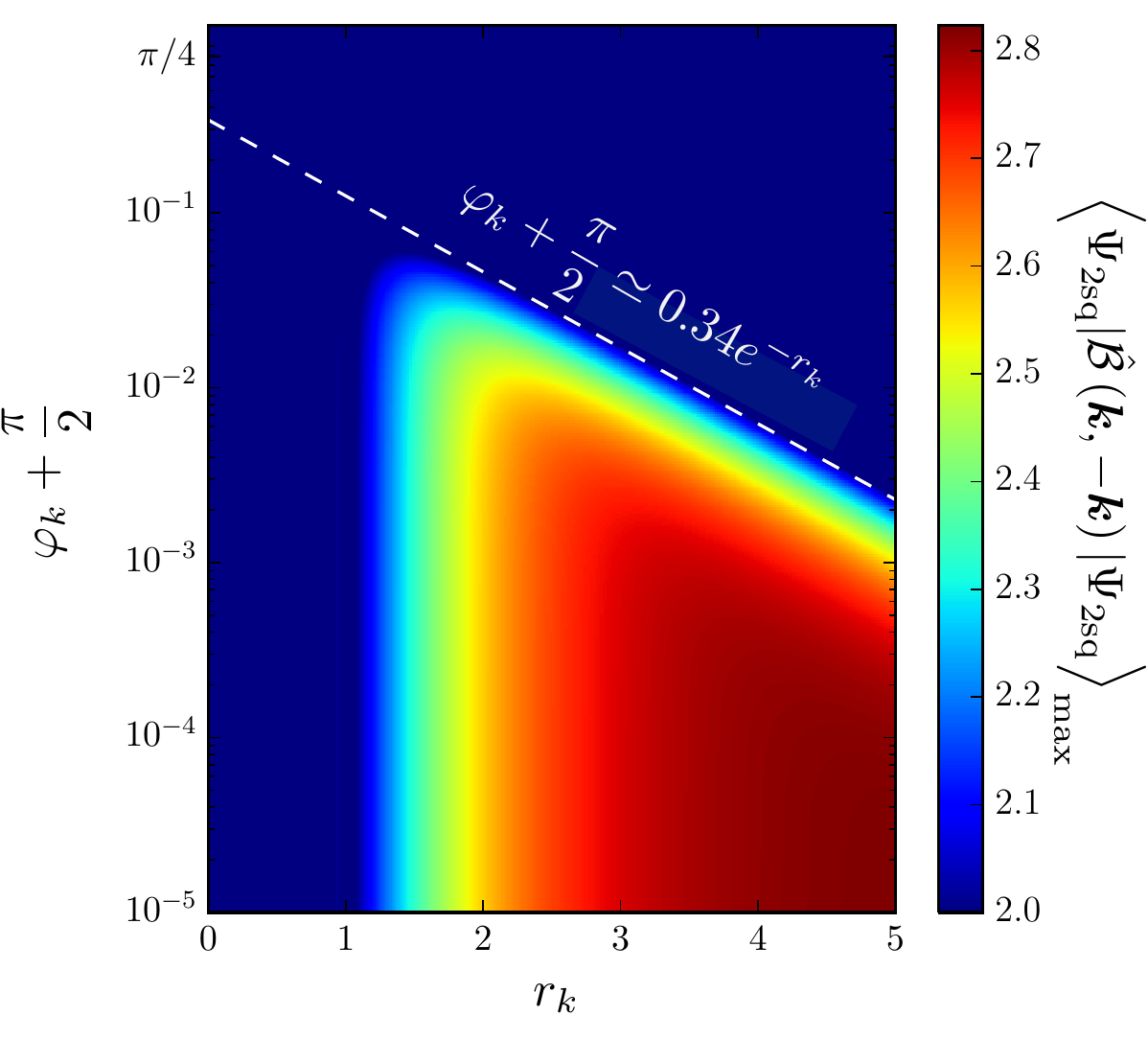}
\end{center}
\caption{Left panel: Mean value
  $\langle \Psi_{2\, {\rm sq}} \vert \hat{{\cal B}}
  \left({\bm k},-{\bm k}\right)\vert \Psi_{2\, {\rm sq}}\rangle$
  of the Bell operators for the BW (solid lines, see \Sec{subsec:bellfakespin}) and GKMR (dashed lines, see \Sec{subsec:alternative}) pseudo-spin operators, when the system is placed in the two-mode squeezed
  state~(\ref{eq:qstate}), as a function of the squeezing parameter
  $r_k$, for different values of the squeezing angle $\varphi_k$. The
  blue lines correspond to $\varphi_k=-\pi/4$, the green lines to
  $\varphi_k=-\pi/3$ and the red lines to $\varphi=-\pi/2$. One can check in
  \Eqs{eq:Bmeanopt} and~(\ref{eq:Bmeanalternative}) that $\varphi_k=-\pi/4$ (blue line) indeed
  leads to a constant
  $\langle \Psi_{2\, {\rm sq}} \vert \hat{{\cal B}}
  \left({\bm k},-{\bm k}\right)\vert \Psi_{2\, {\rm sq}}\rangle=2$.
  The horizontal black line represents the Cirel'son
  bound~\cite{1980LMaPh...4...93C}. Right panel: Maximum Bell operator expectation value $\langle \Psi_{2\, {\rm sq}} \vert \hat{{\cal B}}
  \left({\bm k},-{\bm k}\right)\vert \Psi_{2\, {\rm sq}}\rangle$ for the Larsson pseudo-spin operators, see \Sec{subsec:measurablespin}, where extremisation has been performed over $\ell$, as a function of the squeezing parameters $r_k$ and $\varphi_k$. The dashed white line stands for $\varphi_k+\pi/2 =0.34 e^{-r_k}$ which delimits the Bell inequality violation domain in the large-squeezing limit. This figure is adapted from \Ref{Martin:2016tbd}.}
\label{fig:bell}
\end{figure*}
The previous expression clearly demonstrates that the Bell inequality
is violated for any non-vanishing value of $r_k$ if $\varphi_k \neq \pi/4\pm \pi/2$ (with this
configuration) as can be seen in the left panel of
Fig.~\ref{fig:bell}. So we have the confirmation that, at least at
this level of the analysis, the CMB is an interesting playground to
observe a cosmic violation of Bell inequality. Moreover, on super-Hubble scales, one has $r_k\rightarrow \infty $ and
$\varphi_k\rightarrow -\pi/2$. In this limit one has
$\langle \Psi_{2\, {\rm sq}}\vert \hat{{\cal B}}_{_{\rm BW}}
\left({\bm k},-{\bm k}\right)\vert \Psi_{2\, {\rm sq}}\rangle
\rightarrow 2\sqrt{2}$.
This number is the so-called Cirel'son
bound~\cite{1980LMaPh...4...93C} and represents the maximal value that
the Bell operator can take in quantum mechanics. We therefore conclude
that the CMB is placed in a quantum state that maximally violates the
Bell inequality! In fact, this result is not so surprising since we
have seen that the CMB quantum state has a very large quantum
discord. It can even be shown that, in the limit
$r_k \rightarrow +\infty$, the two-mode squeezed state tends towards
the Einstein-Podolski-Rosen (EPR) state~\cite{Einstein:1935rr}.
\subsection{Gour-Khanna-Mann-Revzen (GKMR) spin operators}
\label{subsec:alternative}
Given the result of the previous subsection, a natural question is
whether the choice of the pseudo-spin operators is unique. In fact it
is not. An alternative set of pseudo-spin operators has also been
considered By Gour, Khanna, Mann and Revzen in
\Refs{2004PhLA..324..415G,2005PhRvA..71b2103R} and it is
interesting to discuss how they can be used in a cosmological context
and, of course, if they lead to a Bell inequality violation. Let us
first introduce $\vert {\cal E}_{\bm k}\rangle $ and
$\vert {\cal O}_{\bm k}\rangle$ by
\begin{align}
\vert {\cal E}_{\bm k}\rangle &=\frac{1}{\sqrt{2}}\left(\vert q_{\bm k}\rangle 
+\vert -q_{\bm k}\rangle \right), \\
\vert {\cal O}_{\bm k}\rangle &=\frac{1}{\sqrt{2}}\left(\vert q_{\bm k}\rangle 
-\vert -q_{\bm k}\rangle \right),
\end{align}
where we remind that $\hat{q}_{\bm k}$ is the operator playing the
role of position in the subspace ${\cal E}_{\bm k}$ (recall that the
operator $\hat{v}_{\bm k}$ mixes the modes ${\bm k}$ and $-{\bm
  k}$). Then, one can define the following operators
\begin{align}
\label{eq:defSxalternative}
\hat{\cal S}_x &=\int _0^{+\infty} {\rm d}q_{\bm k}
\left(\vert {\cal E}_{\bm k}\rangle \langle {\cal O}_{\bm k}\vert
+\vert {\cal O}_{\bm k}\rangle \langle {\cal E}_{\bm k}\vert \right),\\
\label{eq:defSyalternative}
\hat{\cal S}_y &=i\int _0^{+\infty} {\rm d}q_{\bm k}
\left(\vert {\cal O}_{\bm k}\rangle \langle {\cal E}_{\bm k}\vert
-\vert {\cal E}_{\bm k}\rangle \langle {\cal O}_{\bm k}\vert \right),\\
\label{eq:defSzalternative}
\hat{\cal S}_z &=-\int _0^{+\infty} {\rm d}q_{\bm k}
\left(\vert {\cal E}_{\bm k}\rangle \langle {\cal E}_{\bm k}\vert
-\vert {\cal O}_{\bm k}\rangle \langle {\cal O}_{\bm k}\vert \right),
\end{align}
and it is easy to see that, as the notations suggest, they satisfy
all the properties required to be the three components of a spin. In
fact, one can show that the operator of
\Eq{eq:defSzalternative}, which can also be written as
$\hat{\cal S}_z=-\int_{-\infty}^{\infty}{\rm d}q_{\bm k}\vert q_{\bm
  k}\rangle \langle -q_{\bm k}\vert $,
is in fact equal to that of \Eq{eq:defSz1}, $\hat{s}_z=\hat{\cal{S}}_z$. Indeed, it is
straightforward to show that the matrix element
$\langle m\vert \hat{s}_z\vert m'\rangle $, where $\hat{s}_z$ is given
by \Eq{eq:defSz1}, is equal to $\pm \delta_{mm'}$ with a plus
sign if $m$ is odd and a minus sign if $m$ is even. On the other hand,
for the operator~(\ref{eq:defSzalternative}), one has
\begin{align}
& \langle m\vert \hat{\cal S}_z\vert m'\rangle
=-\int_{-\infty}^{+\infty}{\rm d}q_{\bm k}\langle m\vert 
q_{\bm k}\rangle \langle -q_{\bm k}\vert m'\rangle 
\nonumber \\ &
=\frac{-(-1)^{m'}}{\sqrt{\pi 2^{m+m'}m!m'!}}
\int_{-\infty}^{+\infty}{\rm d}q_{\bm k}H_m(q_{\bm k})
H_{m'}(q_{\bm k})e^{-q_{\bm k}^2} 
\nonumber \\ &
=-(-1)^m\delta _{mm'},
\end{align}
where $H_m(.)$ is a Hermite polynomial of order
$n$~\cite{Gradshteyn:1965aa}. This result coincides with the result
obtained before for the operator~(\ref{eq:defSz1}). Notice also that
$\hat{\cal S}_x$ and $\hat{\cal S}_y$ can be written as
$\hat{\cal S}_x=\int_0^{\infty}{\rm d}q_{\bm k}\left(\vert q_{\bm
    k}\rangle \langle q_{\bm k}\vert -\vert -q_{\bm k}\rangle \langle
  -q_{\bm k}\vert \right)$
and
$\hat{\cal S}_y=-\int_0^{\infty}{\rm d}q_{\bm k}\left(\vert q_{\bm
    k}\rangle \langle -q_{\bm k}\vert -\vert -q_{\bm k}\rangle \langle
  q_{\bm k}\vert \right)$.

Then, one can proceed exactly as before, namely introduce a Bell
operator and perform the orientational optimization. One then obtains \Eq{eq:Bmeanopt},
with $\langle \Psi_{2\, {\rm sq}} \vert \hat{\cal S}_z({\bm k})\otimes 
\hat{\cal S}_z(-{\bm k})
\vert \Psi_{2\, {\rm sq}}\rangle = 1 $ and the calculation of appendix~\ref{sec:correlSx}, see
\Eq{eq:correlsxappendix}, shows that
\begin{align}
\label{eq:Bmeanalternative}
& \langle \Psi_{2,{\rm sq}}\vert 
\hat{\cal{S}}_x({\bm k})\hat{\cal{S}}_x(-{\bm k})\vert \Psi_{2, {\rm sq}}\rangle 
= \frac{2}{\pi} \nonumber \\ & \times
\arctan\left[\frac{2\tanh(r_k)\cos(2\varphi_k)}
{\sqrt{\tanh^4(r_k)
-2\tanh^2(r_k)\cos(4\varphi_k)+1}}\right].
\end{align}
This gives the exact dependence of this violation with the squeezing
parameters. In the case where $\varphi_k\neq 0$, this expression is
new. The mean value of the Bell operator as a function of the
squeezing parameter, for different values of the squeezing angle, is
represented in the left panel of Fig.~\ref{fig:bell}. The violation is slightly less strong than with the BW pseudo-spin operators introduced in \Sec{subsec:bellfakespin} but clearly, we
have designed another possible situation that leads to a violation of
the Bell inequality. On super-Hubble scales in particular, since $r_k \rightarrow \infty$ and $\varphi_k \rightarrow -\pi/2$, one has $\langle \Psi_{2\, {\rm sq}}\vert \hat{{\cal B}}_{_{\rm GKMR}}
\left({\bm k},-{\bm k}\right)\vert \Psi_{2\, {\rm sq}}\rangle
\rightarrow 2\sqrt{2}$ and the Cirel'son bound is again saturated.

\subsection{Larsson spin operators}
\label{subsec:measurablespin}

So far, we have introduced two different sets of pseudo-spin
operators, and a third one can be defined in the following way~\cite{2004PhRvA..70b2102L,Martin:2016tbd}. The idea is
to divide the real axis in an infinite number of cells
$\left[n\ell, (n+1)\ell\right]$ of length $\ell$, where $n$ is an
integer number running from $-\infty $ to $+\infty$. The
``coarse-grained'' parameter $\ell$ is chosen by the observer. Then one
introduces the following operator
\begin{align}
\label{eq:defsz}
\hat{S}_z(\ell)
& =\sum_{n=-\infty}^{\infty}(-1)^n\int _{n\ell}
^{(n+1)\ell}{\rm d}q_{\bm k} \vert q_{\bm k}\rangle 
\langle q_{\bm k}\vert \, .
\end{align}
This defines a spin variable because the eigenvalues of this operator
are $\pm 1$.

The other pseudo-spin components can be defined by means of the
following operators
\begin{align}
\label{eq:defsx}
\hat{S}_x(\ell)& =\hat{S}_+(\ell)+\hat{S}_-(\ell), \\
\label{eq:defsy}
\hat{S}_y(\ell) & =-i\left[\hat{S}_+(\ell)-\hat{S}_-(\ell)\right], 
\end{align}
with 
\begin{align}
  \hat{S}_+(\ell)
  =\sum_{n=-\infty}^{\infty}\int_{2n\ell}^{(2n+1)\ell}{\rm
  d}q_{\bm k}\left\vert q_{\bm k}\rangle \langle q_{\bm k}
+\ell\right\vert
\label{eq:defsplus}
\end{align}
and $\hat{S}_-(\ell)=\hat{S}_+^{\dagger}(\ell)$. It can be easily
shown~\cite{2004PhRvA..70b2102L,Martin:2016tbd} that $\hat{S}_x$,
$\hat{S}_y$ and $\hat{S}_z$ satisfy all the properties of a spin
operators system. It is then sufficient to proceed as before, namely to
define a pseudo Bell operator using the spin operators that we have
just introduced, perform the orientational optimization and, then,
study its mean value in a two-mode squeezed state. In the case of the
operators~(\ref{eq:defsz}), (\ref{eq:defsx}) and~(\ref{eq:defsy}),
there does not exist a simple, explicit, analytical expression for the
spin correlators, similar to \Eqs{eq:sxsx:BW} and~(\ref{eq:Bmeanalternative}).
The calculation has therefore to be done
numerically. It was carried out in \Ref{Martin:2016tbd} and, in
that paper, it was shown that the Bell inequality is indeed violated
in that case. In the right panel of \Fig{fig:bell}, we have reproduced from \Ref{Martin:2016tbd} a map of the maximal Bell operator expectation value (where the extremisation is performed over $\ell$) as a function of the squeezing parameters. We thus have a third example of a design that violates the Bell inequality.

Contrary to the two previous examples however, in the super-Hubble limit, the Cirel'son
bound is not necessarily saturated, and the exact value of the Bell operator expectation value depends on the details of the inflationary dynamics. In the large-squeezing limit indeed, in \Ref{Martin:2016tbd}, it was shown that all pseudo-spin correlation functions depend only on the combination (see footnote~\ref{footnote:varphi:convention}) $(\varphi_k+\pi/2) e^{r_k}$, and Bell inequality violation is obtained when $(\varphi_k+\pi/2)e^{r_k}<0.34$ (see the right panel of \Fig{fig:bell} and/or Fig. 16 of \Ref{Martin:2016tbd}). In the super-Hubble limit, $\varphi_k+\pi/2 \rightarrow 0$ but $e^{r_k} \rightarrow \infty$ so a more detailed analysis is required. At leading order in the slow-roll approximation, the Mukhanov-Sasaki variable takes the form $v_{\bm k}(\eta) = \sqrt{\pi/k}/2 \sqrt{-k \eta} e^{-i\pi(\nu+1/2)/2} H_\nu^{(2)}(-k\eta)$ where we recall that $\eta$ is the conformal time, $H_\nu^{(2)}$ is the Hankel function of the second kind, and $\nu$ is a constant that can be related to the spectral index $\nS$ of the curvature perturbations power spectrum according to $\nu=3/2+(\nS-1)/2$, where this expression is valid again at leading order in slow roll. The relationship between the squeezing parameters and the Mukhanov-Sasaki variable can be obtained by combining Eqs.~(12), (16), (24) and~(25) of \Ref{Martin:2015qta} and one obtains
\begin{widetext}
\bea
\label{eq:coshr:v}
\cosh^2 r_k & = & \dfrac{1}{8 k v_{\bm k} v^*_{\bm k}} \left[ \left(v^*_{\bm k} v'_{\bm k} + v_{\bm k} v'^*_{\bm k}-2\dfrac{z'}{z}v_{\bm k} v^*_{\bm k}\right)^2 + \left(1+2k v_{\bm k} v_{\bm k}^* \right)^2\right]\, ,
\\ 
\label{eq:tanphi:v}
\tan \left(2\varphi_k\right) & = &4 k v_{\bm k} v_{\bm k}^*\dfrac{v_{\bm k}^* v_{\bm k}' + v_{\bm k} v_{\bm k}'^*-2\dfrac{z'}{z}v_{\bm k} v_{\bm k}^*}{\left(2kv_{\bm k}v_{\bm k}^*\right)^2 -1 -\left(v_{\bm k}^* v_{\bm k}' + v_{\bm k} v_{\bm k}'^*-2\dfrac{z'}{z}v_{\bm k} v_{\bm k}^*\right)^2}\, .
\eea
\end{widetext}
Using the fact that, in slow-roll inflation, $z'/z\simeq (1/2-\nu)/\eta$, in terms of the number of \efolds $N-N_*(k)$ spent by the mode $k$ outside the Hubble radius, one obtains, in the super-Hubble limit and at leading order in slow roll,
\bea
\label{eq:Larsson:magicalComb}
\left(\varphi_k+\frac{\pi}{2}\right) e^{r_k} \simeq \exp\left\lbrace{\frac{1-\nS}{2}\left[N- N_*(k)\right]}\right\rbrace\, .
\eea
In the case of de Sitter ($\nS=1$), this combination turns out to be time independent and equal to one, meaning no Bell inequality violation. In the more general case, however, this is clearly a time-dependent quantity. For $(\varphi_k+\pi/2)e^{r_k}$ to be smaller than $0.34$, one can see that the power spectrum has to be blue, $\nS>1$, which is excluded by the data~\cite{Ade:2015xua}. Therefore, unless slow roll is violated at some point during inflation, the Larsson pseudo-spin operators are not the best candidates to yield a Bell inequalities violation. 

To our knowledge, no other pseudo-spin operators for continuous variables have
been proposed and, therefore, we have now covered all the cases. In
the next section, we discuss these results and consider the question
of whether these spin operators can really be measured on the sky.

\section{Discussion}
\label{sec:discussion}

\subsection{Bell inequality violation and the positivity 
of the Wigner function}
\label{eq:bivwigner}

Let us start by a remark about the consistency of the previous
results where Bell inequality violations are obtained with a non-negative Wigner function. We notice that some of the pseudo-spin operators we have introduced are improper dynamical variables, a notion that we now
explain with an example (a more complete discussion can be found in
\Ref{2005PhRvA..71b2103R}). Let
$\hat{A}={\cal A}(\hat{q}_{\bm k},\hat{p}_{\bm k})$ be a general
operator (we restrict ourselves to a one-dimensional phase space but the
generalization to higher dimensional phase spaces is trivial), where we
use the position and momentum introduced before. Its Wigner-Weyl
representation is defined by
\begin{align}
W_{\hat{A}}(q_{\bm k},\pi_{\bm k}) \equiv &
\int _{\setR} {\rm d}x \, e^{-i\pi_{\bm k} x}
\left\langle q+\frac{x}{2}\biggl \vert
{\cal A}\left(\hat{q}_{\bm k},\hat{p}_{\bm k}\right)
\biggr \vert q_{\bm k}-\frac{x}{2}
\right \rangle.
\end{align}
This is clearly a function (\ie not an operator) on phase space. Let
us now consider a (bounded) function ${\cal F}(.)$ defined on the real axis $\mathbb{R}$. We can then introduce the following
operator
\begin{align}
\hat{F}\equiv \int _{-\infty}^{+\infty}{\rm d}q'_{\bm k}
\vert q'_{\bm k}\rangle {\cal F}(q'_{\bm k})\langle 
q'_{\bm k}\vert .
\end{align}
The eigenvalues of the operator $\hat{F}$ are precisely ${\cal F}(\mathbb{R})$
(this is a way to define an operator from its spectrum). Then, it is a
trivial calculation to show that
\begin{align}
W_{\hat{F}}(q_{\bm k},\pi_{\bm k}) &={\cal F}(q_{\bm k}).
\end{align}
So the function $W_{\hat{F}}(q,\pi)$ takes all and only the eigenvalues
of the operator $\hat{F}$. We then say that $\hat{F}$ is a proper
dynamical variable. Of course, the previous calculation is only an
illustrative example and can easily be generalized, for instance to
functions of $\pi_{\bm k}$, see \Ref{2005PhRvA..71b2103R}.

Now, it is easy to show that the pseudo-spin operators that we have
introduced above are not proper dynamical variables. For instance, the
Wigner-Weyl representative of the BW operator $\hat{s}_z$ given by
\Eq{eq:defSz1} can be expressed as
\begin{align}
W_{\hat{s}_z}(q_{\bm k},\pi_{\bm k})&\equiv 
\int _{-\infty}^{+\infty}e^{-i\pi_{\bm k} x}{\rm d}x
\left\langle q_{\bm k}+\frac{x}{2}\left \vert
\hat{s}_{z}\right \vert q_{\bm k}-\frac{x}{2}
\right \rangle \nonumber \\ &
=-\pi \delta(q_{\bm k})\delta(\pi_{\bm k}).
\end{align}
Clearly, the function $W_{\hat{s}_z}(q_{\bm k},\pi_{\bm k})$ does not
consist of only two values $\pm 1$.  

For the GKMR spin operators defined by
\Eqs{eq:defSxalternative}, (\ref{eq:defSyalternative})
and~(\ref{eq:defSzalternative}), we reach the same conclusions, since we have
shown that the GKMR operator~(\ref{eq:defSzalternative}) is in fact
the same as the BW operator~(\ref{eq:defSz1}). The Wigner-Weyl representation of
$\hat{\cal S}_x$ given by \Eq{eq:defSxalternative} can be
written as
\begin{align}
\label{eq:weylsx}
  W_{\hat{\cal S}_x}(q_{\bm k},\pi_{\bm k})=\mathrm{sign}(q_{\bm k}),
\end{align}
\ie it is a proper dynamical variable according to the previous
definition. Finally, the Wigner-Weyl representation of the
GKMR $y$-component~(\ref{eq:defSyalternative}) is
$W_{\hat{\cal S}_y}(q_{\bm k},\pi_{\bm k})=-\delta(q_{\bm k}){\cal
  P}(1/\pi_{\bm k})$,
where ${\cal P}$ denotes the principal value and therefore the
operator~(\ref{eq:defSyalternative}) is not proper.

It is also interesting to calculate the Wigner-Weyl transform of the
Larsson pseudo-spin operators~(\ref{eq:defsx}), (\ref{eq:defsy})
and~(\ref{eq:defsz}). Straightforward manipulations lead to
\begin{widetext} 
\begin{align}
W_{\hat{S}_x}(q_{\bm k},\pi_{\bm k}) &=2\sum_{n=-\infty}^{+\infty}
\cos\left(\pi_{\bm k} \ell\right)
\biggl[\Theta \left(q_{\bm k}-n \ell-\frac{\ell}{2}\right)
-\Theta\left(q_{\bm k}-n\ell-\frac32\ell\right)\biggr],
\nonumber \\ 
W_{\hat{S}_y}(q_{\bm k},\pi_{\bm k}) &=2\sum_{n=-\infty}^{+\infty}
\sin\left(\pi_{\bm k} \ell\right)
\left[\Theta \left(q_{\bm k}-n \ell-\frac{\ell}{2}\right)
-\Theta\left(q_{\bm k}-n\ell-\frac32\ell\right)\right],
\nonumber \\
W_{\hat{S}_z}(q_{\bm k},\pi_{\bm k}) &=\sum_{n=-\infty}^{+\infty}(-1)^n
\left[\Theta \left(q_{\bm k}-n \ell\right)
-\Theta\left(q_{\bm k}-n\ell-\ell\right)\right],
\end{align}
\end{widetext}
where $\Theta(.)$ is the Heaviside function. The above equations show
that $\hat{S}_z$ is a proper variable but $\hat{S}_x$ and $\hat{S}_y$
are not.

We conclude that each of the three sets of pseudo-spin operators
introduced before contains, at least, one improper variable. And this
makes perfect sense since there is a theorem stating that, if the
Wigner function is positive definite, a violation of the Bell
inequality can only occur for improper dynamical variables, see
\Ref{2005PhRvA..71b2103R} for a more accurate discussion of this
point. This is clearly relevant for the CMB case since, as already
mentioned, a two-mode squeezed state has a positive definite Wigner
function. In fact, the CMB is a prototypical situation where this
theorem is useful. 

This however says nothing about about the measurability of the effect, which we now discuss.

\subsection{Are the pseudo-spin operators measurable?}
\label{subsec:spinmeasurability}

Answering the question asked in the title of this subsection is not an
easy task since it involves the measurement problem of quantum
mechanics in the context of cosmology. It seems however reasonable to assume
that the temperature anisotropy operator,
\begin{align}
\widehat{\frac{\delta T}{T}}(\theta , \phi)
=\sum _{\ell =2}^{+\infty}\sum _{m=-\ell}^{m=\ell}\hat{a}_{\ell m}Y_{\ell m}
(\theta,\phi),
\end{align}
is an observable, since it is a real quantity. More precisely, it is a family of operators
parametrized by the continuous labels $\theta $ and $\phi$. In the above expression, the
coefficients $\hat{a}_{\ell m}$ are a collection of, non-hermitian,
operators, as needed for the consistency of this equation. Let us now
try to relate the temperature fluctuation operator to the curvature
operator $\hat{\zeta}_{\bm k}$ introduced before. The Sachs-Wolfe
effect implies that
\begin{align}
\label{eq:sweffect}
\frac{\delta T}{T}({\bm e})
=& \int \frac{{\rm d}{\bm k}}{(2\pi)^{3/2}}
\left[F({\bm k})+i{\bm k}\cdot {\bm e}\, G({\bm k})\right]
\nonumber \\ & \times 
e^{-i {\bm k}\cdot 
{\bm e}(\eta_{\rm lss}-\eta_0)
+i{\bm k}\cdot {\bm x}_{0}}\, ,
\end{align}
where ${\bm e}$ is a unit vector in the direction labeled by the
angles $\theta$ and $\phi$. The quantities $\eta_{\rm lss}$ and
$\eta_0$ are the last scattering surface (lss) and present day
($0$) conformal times while ${\bm x}_0$ represents Earth's
location. The functions $F({\bm k})$ and $G({\bm k})$ are the form
factors and describe the evolution of the perturbation in the
post-inflationary universe. The important property is that these form
factors are proportional to $\zeta_{\bm k}(\eta_{\rm end})$ evaluated
at the end of inflation. Strictly speaking, they also depend on the
derivative of $\zeta_{\bm k}$ but the point is that this dependence is
completely negligible since it is related to the presence of a
decaying mode. However, in principle, $\zeta_{\bm k}'$ is
present. This means that, at the operator level, one can write
\begin{align}
\widehat{\frac{\delta T}{T}}({\bm e})
=& \int \frac{{\rm d}{\bm k}}{(2\pi)^{3/2}}
\left[F({\bm k})+i{\bm k}\cdot {\bm e}\, G({\bm k})\right]
\hat{\zeta}_{\bm k}(\eta_{\rm end})
\nonumber \\ & \times 
e^{-i {\bm k}\cdot 
{\bm e}(\eta_{\rm lss}-\eta_0)
+i{\bm k}\cdot {\bm x}_{0}}\, ,
\end{align}
where, compared to \Eq{eq:sweffect}, we have slightly redefined
the form factors. In particular, the above expression implies that
$\widehat{\delta T/T}$ for two different directions $\bm{e}$ and $\bm{e}'$ are commuting operators
since $[\hat{\zeta}_{\bm k},\hat{\zeta}_{\bm p}]=0$. Notice that this
result crucially rests on the fact that we have neglected the decaying
mode. If not, the temperature anisotropy would depend on
$\zeta_{\bm k}'$, namely on the momentum and, therefore, it would no
longer commute for different directions on the sky. We also conclude
that, the eigenvectors of $\widehat{\delta T/T}(\theta,\phi)$, that we
denote $\ketmap$, are those of $\hat{\zeta}_{\bm k}+\hat{\zeta}^\dagger_{\bm k}$.
Then, given the fact that
$\hat{\zeta}_{\bm k}=(c_{\bm k}+c_{-{\bm k}}^{\dagger})/\sqrt{2k}$,
one can check that the two-mode squeezed state introduced before
is not an eigenstate of the temperature anisotropy operator, which
means that
\begin{align}
\vert \Psi_{2\, {\rm sq}}\rangle =\sum_{\map}c(\map) \ketmap\, .
\end{align}
Since we observe a specific map, one has then to assume that, after
our ``observation'' of the sky, the system is placed in the
``eigenstate'' $\ketmapPlanck_{\rm Planck} $ corresponding to
the ``eigenvalue'' $\delta T/T(\theta,\phi)\vert_{\rm Planck}$, namely
the Planck map. How the process
\begin{align}
\vert \Psi_{2\, {\rm sq}}\rangle=
\sum_{ \map}c( \map)  \ketmap \rightarrow 
\ketmapPlanck_{\rm Planck} ,
\end{align}
occurred is of course the quantum measurement problem which is usually
``solved'' by the collapse postulate. However, the status of this
postulate is, to say the least, unclear in cosmology and alternatives
to the standard Copenhagen interpretation have been proposed such as,
Continuous Spontaneous Localization (CSL)
models~\cite{Perez:2005gh,Sudarsky:2009za,Bassi:2012bg,Martin:2012pea,
  Canate:2012ua,Das:2013qwa,Das:2014ada,Leon:2015hwa}, the
manyworlds~\cite{Mukhanov:2007zza,Kiefer:2008ku} or the Bohm-de
Broglie interpretations~\cite{Valentini:2008dq,PintoNeto:2011ui,
  Goldstein:2015mha,Colin:2015tla}. Here, we will not attempt to
discuss these issues and will just assume that the system is placed in
a specific eigenvector corresponding to a specific eigenvalue, namely
the Planck map  $\ketmapPlanck_{\rm Planck} $.

Let us now see what it means to ``measure'' the spin operators.
According to the previous discussion, it seems reasonable to assume
that we have measured $\hat{\zeta}_{\bm k}$, \ie we have a collection
of numbers $\zeta_{\bm k}$. Of course, one can only measure real
quantities and, therefore, one should rather say that we have measured
the real and imaginary parts of the operators $\hat{\zeta}_{\bm k}$.
But this is equivalent since all these quantities commute and,
therefore, a measurement of the real and imaginary parts of the
curvature perturbations is also a measurement of $\hat{\zeta}_{\bm k}$
and its hermitian conjugate, exactly as a measurement of
$\hat{\cal O}$ is also a measurement of any function $f(\hat{\cal O})$
since $[\hat{\cal O},f(\hat{\cal O})]=0$ [if we have found
$\hat{\cal O}$ to be the number $o$, then, because of their vanishing
commutator, we are entitled to say that we have measured
$f(\hat{\cal O})$ to be $f(o)$]. Then, using \Eqs{eq:linkvq}
and~(\ref{eq:linkpq}), it is easy to establish that
\begin{align}
\hat{q}_{\bm k}=\frac{z}{2}\left(\hat{\zeta}_{\bm k}
+\hat{\zeta}_{-{\bm k}}\right)
+\frac{z}{2k}\left(\zeta_{\bm k}'-\zeta_{-{\bm k}}'\right),
\end{align}
and we see that the knowledge of $\hat{\zeta}_{\bm k}$ is not
sufficient to infer $\hat{q}_{\bm k}$. However, if the decaying mode
is neglected, then a measurement of $\hat{\zeta}_{\bm k}$ is a
measurement of the operator $\hat{q}_{\bm k}$. Notice that in this limit, $\hat{q}_{\bm k} = \hat{q}_{- \bm{k}}$, hence $\hat{\mathcal{S}}_x({\bm k}) = \hat{\mathcal{S}}_x( -{\bm k})$ [since they share the same Wigner-Weyl representation according to \Eq{eq:weylsx}], thus $\hat{\mathcal{S}}_x({\bm k}) \otimes \hat{\mathcal{S}}_x( -{\bm k})= \hat{{\mathrm{I}}}$. As a consequence, $\langle \Psi_{2\mathrm{sq}} \vert \hat{\mathcal{S}}_x({\bm k}) \otimes \hat{\mathcal{S}}_x( -{\bm k}) \vert   \Psi_{2\mathrm{sq}} \rangle = 1$, which is consistent with the large-squeezing limit of \Eq{eq:Bmeanalternative}. Moreover, since  $\langle \Psi_{2\mathrm{sq}} \vert \hat{\mathcal{S}}_z({\bm k}) \otimes \hat{\mathcal{S}}_z( -{\bm k}) \vert   \Psi_{2\mathrm{sq}} \rangle $ is always $1$, this also shows, as announced in the introduction, that a Bell inequality violation (furthermore a maximal one) can be obtained even when the decaying mode is neglected.

In the following, we
assume that the decaying mode can be ignored and study the consequences. The question
is now, given the knowledge of the numbers $q_{\bm k}$, can we infer
the values of the pseudo-spin operators?
Let us first discuss the BW spin operators defined by
\Eqs{eq:defSx1}, (\ref{eq:defSy1}) and~(\ref{eq:defSz1}).  One
can show that
\begin{equation}
\hat{s}_z({\bm k})=-(-1)^{\hat{N}_{\bm k}}=-\int_{-\infty}^{+\infty}{\rm d}q_{\bm k}
\vert q_{\bm k}\rangle \langle -q_{\bm k}\vert,
\end{equation}
where we recall that
$\hat{N}_{\bm k}=\hat{c}_{\bm k}^{\dagger}\hat{c}_{\bm k}$ is the
particle number operator. This operator does not commute with
$\hat{q}_{\bm k}$ since one has 
\bea
  \langle q_{\bm k}\vert \left[\hat{s}_z({\bm k}),\hat{q}_{\bm k}\right]
  \vert q_{\bm k}'\rangle &=\left(q_{\bm k}-q_{\bm k}'\right)
  \langle -q_{\bm k}\vert q_{\bm k}'\rangle\nonumber \\
 & =2  q_{\bm k} \delta(  q_{\bm k}+  q_{\bm k}^\prime)
  \neq 0.
\eea
Since, in cosmology, we are given a measurement and, contrary to what
happens in the laboratory, we cannot perform a new measurement, this
means that one simply cannot measure $\hat{s}_z(\bm{k})$ if $\hat{q}_{\bm k}$ has been measured. On the other hand,
the operators $\hat{s}_x$ and $\hat{s}_y$ are related to the parity
flip operators $\hat{s}_{\pm}$ through
$\hat{s}_{\pm}\equiv (\hat{s}_x\pm i\hat{s}_y)/2$, where
\begin{equation}
\label{eq:sminus}
\hat{s}_-({\bm k})=\left[{\hat{\mathrm{I}}}+\left(-1\right)^{\hat{N}_{\bm k}}
\right]\frac{1}{2\sqrt{\hat{N}_{\bm k}+1}}\hat{c}_{\bm k},
\end{equation}
and $\hat{s}_+=\hat{s}_-^{\dagger}$. It is not obvious to design an
experimental protocol in order to measure these operators. It is not
even clear whether this is, in principle, feasible. In any case,
since the operator $\hat{c}_{\bm k}$ depends on $\hat{q}_{\bm k}$ and
$\hat{\pi}_{\bm k}$, it is hard to see how $\hat{s}_{\pm}$ could
commute with $\hat{q}_{\bm k}$. So, as far as the BM spins defined by
\Eqs{eq:defSx1}, (\ref{eq:defSy1}) and~(\ref{eq:defSz1}) are
concerned, we are in a situation where it is probably impossible to
infer their values from the data.

The same conclusion is also valid for the GKMR operators defined in
\Eqs{eq:defSxalternative}, (\ref{eq:defSyalternative})
and~(\ref{eq:defSzalternative}) since, for instance, they share the
same $z$-component spin operator as the one of the BW proposal, namely $\hat{s}_z=\hat{\cal S}_z$.
However, it is interesting to notice that 
\begin{align}
  \left\langle q_{\bm k}\left\vert \left[\hat{\cal S}_x({\bm k}),\hat{q}_{\bm k}\right]
  \right\vert q_{\bm k}'\right\rangle =0,
\end{align}
which means that $\hat{\cal S}_x({\bm k})$ can be measured. In fact, it 
is also easy to show that
\begin{align}
\hat{\cal S}_x\vert q_{\bm k}\rangle ={\mathrm{sign}}(q_{\bm k})\vert q_{\bm k}\rangle,
\end{align}
which means that, once we are given the number $q_{\bm k}$, the value
of $\hat{\cal S}_x$ is just the sign of $q_{\bm k}$. Of course, this
is compatible with its Weyl-Wigner representation, see
\Eq{eq:weylsx}. The fact that $\hat{\cal S}_x({\bm k})$ can be
measured does not ``save'' the GKMR operators since we would need
another measurable operator, which is not the case for the only one
left, namely $\hat{\cal S}_y$, since 
\bea
  \langle q_{\bm k}\vert \left[\hat{\cal S}_y({\bm k}),\hat{q}_{\bm k}\right]
  \vert q_{\bm k}'\rangle &=\left(q_{\bm k}-q'_{\bm k}\right)\langle -q_{\bm k}
\vert q'_{\bm k}\rangle+q'_{\bm k}\langle q_{\bm k}
\vert -q'_{\bm k}\rangle
\nonumber \\ & +q_{\bm k}\langle q_{\bm k}
\vert q'_{\bm k}\rangle
\nonumber \\ &= q_{\bm k} \left[ \delta\left(q_{\bm k}+q_{\bm k}^\prime\right) + \delta\left(q_{\bm k}-q_{\bm k}^\prime\right)\right]\neq 0\, . \nonumber \\ 
\eea

Finally remains the Larsson operators~(\ref{eq:defsx}),
(\ref{eq:defsy}) and~(\ref{eq:defsz}). In particular, the $\hat{S}_z$
operator is measurable since one has
\begin{align}
  \left\langle q_{\bm k}\left\vert \left[\hat{S}_z({\bm k}),\hat{q}_{\bm k}\right]
  \right\vert q_{\bm k}'\right\rangle =0.
\end{align}
Indeed, it is easy to show that
\begin{align}
\hat{S}_z\vert q_{\bm k}\rangle =
\sum _{n=-\infty}^{n=\infty}(-1)^n\left[\Theta(q_{\bm k}-n\ell)
-\Theta(q_{\bm k}-n\ell-\ell)\right]\vert q_{\bm k}\rangle.
\end{align}
This formula tells us that, in practice, the observer chooses a value
of $\ell$ and, given a measurement of $\hat{q}_{\bm k}$, identifies
the value of $n$ such that $q_{\bm k}\in [n\ell,(n+1)\ell]$. The
measurement of $\hat{S}_z(\ell)$ is then $(-1)^n$. However, neither
$\hat{S}_x(\ell)$ nor $\hat{S}_y(\ell)$ can be inferred from the
knowledge of $q_{\bm k}$ since 
\begin{align}
  \left\langle q_{\bm k}\left\vert \left[\hat{S}_x({\bm k}),\hat{q}_{\bm k}\right]
  \right\vert q_{\bm k}'\right\rangle & \neq 0, \quad
\left\langle q_{\bm k}\left\vert \left[\hat{S}_y({\bm k}),\hat{q}_{\bm k}\right]
  \right\vert q_{\bm k}'\right\rangle \neq 0,
\end{align}
as can be established from the result
\begin{align}
  \left\langle q_{\bm k}\left\vert \left[\hat{S}_+({\bm k}),\hat{q}_{\bm k}\right]
  \right\vert q_{\bm k}'\right\rangle &= 
\ell \sum_{n=-\infty}^{n=+\infty}\biggl[\Theta\left(q_{\bm k}-2n\ell\right)
\nonumber \\ &- \Theta\left(q_{\bm k}-2n\ell-\ell\right)\biggr]
\delta(q_{\bm k}'-q_{\bm k}-\ell), \\
\left\langle q_{\bm k}\left\vert \left[\hat{S}_-({\bm k}),\hat{q}_{\bm k}\right]
  \right\vert q_{\bm k}'\right\rangle &=
-\ell \sum_{n=-\infty}^{n=+\infty}\biggl[\Theta\left(q_{\bm k}-2n\ell-\ell\right)
\nonumber \\ &-
\Theta\left(q_{\bm k}-2n\ell-2\ell\right)\biggr]
\delta(q_{\bm k}'-q_{\bm k}+\ell).
\end{align}

We conclude this section by stressing out that the knowledge of the
numbers $q_{\bm k}$ is not sufficient to determine two spin operators,
which is necessary to observe a Bell inequality violation in the
data. This limitation seems to be deeply rooted in the fact that we
work in a cosmological context. Indeed, in a conventional situation,
one would first measure, say, the $z$-component of the spin and, then,
in a second time, one would repeat the experiment and measure, say,
the $x$-component. In cosmology, one cannot repeat the experiment and,
in some sense, we are given the measurement. This means that we are
unable to determine the value of two non-commuting observables from
the data. In the laboratory, this would be like being given a measure
of the $z$-component only and trying to infer a Bell inequality
violation from this single measurement. A possible way out would be to
use a kind of ergodic theorem. Indeed, as we have already seen,
$\widehat{\delta T/T}(\theta , \phi)$ is in fact a collection of
operators, one for each direction in the sky. One could then imagine to
perform a measurement of the $x$-component in one direction and of
the $y$-component in another direction. However, at least if the
Copenhagen interpretation is taken to its logical extreme, this is
already ``too late'', since we have already measured the operator
$\hat{q}_{\bm k}$ (or $\hat{\zeta}_{\bm k}$) over the entire celestial
sphere. Therefore, at this point, it seems that there is no hope to
observe a Bell inequality violation in the CMB data unless we find a
way to go beyond these fundamental limitations.

\subsection{Decoherence}
\label{sec:decoherence}
\begin{figure}[t]
\begin{center}
\includegraphics[width=8cm]{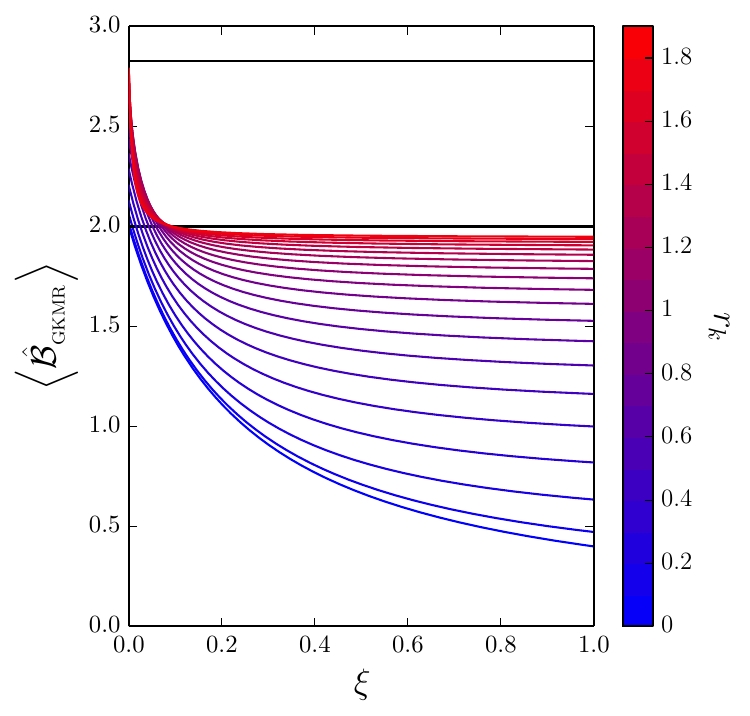}
\end{center}
\caption{Mean value of the Bell operator for the GKMR pseudo-spin operators in presence of decoherence parametrized by the quantity $\xi$. Different values of the squeezing parameter $r_k$ are displayed with different colors, and the squeezing angle is chosen to be $\varphi_k=-\pi/2$.}
\label{fig:decoherence}
\end{figure}
It is also interesting to study how robust our results are against quantum decoherence~\cite{Zurek:1981xq, Zurek:1982ii,
  Joos:1984uk, Paz:1993tg, Calzetta:1995ys, Franco:2011fg}. For this purpose, we model decoherence through the generic Gaussian channel which consists in replacing the density matrix $\rho$ according to~\cite{Joos:1984uk} $\rho(q_{\bm{k}},q_{-\bm{k}},\tilde{q}_{\bm{k}},\tilde{q}_{-\bm{k}})\rightarrow\rho(q_{\bm{k}},q_{-\bm{k}},\tilde{q}_{\bm{k}},\tilde{q}_{-\bm{k}})\exp[-\xi(q_{\bm{k}}-\tilde{q}_{\bm{k}})^2-\xi(q_{-\bm{k}}-\tilde{q}_{-\bm{k}})^2]$,
where the phenomenological parameter $\xi$ encodes the interaction strength with the environment. Given a concrete model for the environement, in principle, one could determine $\xi$ explicitly. We illustrate the effect of decoherence with the GKMR pseudo-spin operators~(\ref{eq:defSxalternative})-(\ref{eq:defSzalternative}). As already mentioned, the mean value of the Bell operator is given by an equation similar to \Eq{eq:Bmeanopt}, where $\langle \Psi_{2\, {\rm sq}} \vert \hat{\cal S}_x({\bm k})\otimes 
\hat{\cal S}_x(-{\bm k})
\vert \Psi_{2\, {\rm sq}}\rangle $ is still given by \Eq{eq:Bmeanalternative}, while, instead of $\langle \Psi_{2\, {\rm sq}} \vert \hat{\cal S}_z({\bm k})\otimes 
\hat{\cal S}_z(-{\bm k})
\vert \Psi_{2\, {\rm sq}}\rangle =1 $, one now has
\begin{align}
& \langle \Psi_{2\, {\rm sq}} \vert \hat{\cal S}_z({\bm k})\otimes 
\hat{\cal S}_z(-{\bm k})
\vert \Psi_{2\, {\rm sq}}\rangle  = 
\left[1+8\xi\cosh\left(2r_k\right)
\right. \nonumber \\  & \left.
+12\xi^2+4\xi^2\cos\left(4\varphi_k\right)+8\xi^2\cosh(4r_k)\sin^2\left(2\varphi_k\right)\right]^{-1/2}\, .
\end{align}
The corresponding Bell operator mean value is shown in \Fig{fig:decoherence}. As expected, one can see that as $\xi$ increases, coherence is lost, the mean value of $\hat{B}_{{}_\mathrm{GKMR}}$ becomes smaller than $2$ and Bell inequality violation no longer exists. For large values of $r_k$, the disappearance of Bell inequality violation occurs rapidly, even though $\langle \hat{B}_{{}_\mathrm{GKMR}} \rangle$ remains close to (while being less than) the Bell violation threshold $2$. This is a consequence of the well-known fragility of highly-squeezed states under environmental influence~\cite{Giulini:1996nw}. This result suggests that decoherence is another challenge for Bell CMB experiments.

\section{Conclusions}
\label{sec:conclusions}

Inflation implies that the quantum state of CMB anisotropies is a
two-mode squeezed state. This state is an entangled state and, as a
consequence, it is natural to imagine that a cosmic version of the
Bell experiment could reveal quantum correlations in the sky. In this
article, we have discussed the problems one faces when one
tries to implement this program concretely. We have exhibited
different possible setups and identified those which, in principle,
could be realized. However, we have also shown that all of them are such
that only one component of the pseudo-spin operators can at most, in practice,
be inferred from the data, which is not sufficient to observe a Bell
inequality violation. In some sense, we encounter again the well-known
problem that the quantum behavior of the perturbations is hidden in the
decaying mode, but more crucially, we also face the fact that one cannot perform repeated measurements 
of the CMB along different spin operators, as is necessary in standard Bell experiments.
The approach followed in this
article attempts to circumvent these problems from
a completely new perspective and, therefore, sheds new light on these
questions. In particular, it shows that the only way out is to design a
method which makes use of only one pseudo-spin component, either $\hat{\cal S}_x$
or $\hat{S}_z$. This turns out to be possible in the Leggett-Garg
inequality~\cite{PhysRevLett.54.857,2014RPPh...77a6001E} proposal where the same spin component is
measured at different times. Since it
was recently shown~\cite{Martin:2016nrr} that the Leggett-Garg inequality is violated if the
state of the system is a squeezed state, this
opens up another possibility for hunting down the quantum origin of the
cosmological structures about which we plan to report soon.
\vspace{0.4cm}
\begin{acknowledgments}
  V.V. acknowledges funding from the European Union's Horizon 2020 research and innovation programme under the Marie Sk\l odowska-Curie grant agreement N${}^0$ 750491
  and financial support from STFC grants ST/K00090X/1
  and ST/N000668/1.
\end{acknowledgments}
\appendix
\section{Correlation function of $\hat{\cal{S}}_x$}
\label{sec:correlSx}

In this appendix, we show how to arrive at
\Eq{eq:Bmeanalternative}. The $\hat{\cal{S}}_x({\bm k})$ operator
defined in \Eq{eq:defSxalternative} can also be written as
\begin{align}
\hat{\cal{S}}_x({\bm k})=\int _0^{+\infty}{\mathrm{d}}\tilde{q}_{\bm k}
\left(\vert \tilde{q}_{\bm k}\rangle 
\langle \tilde{q}_{\bm k}\vert 
-\vert -\tilde{q}_{\bm k}\rangle 
\langle -\tilde{q}_{\bm k}\vert \right).
\end{align}
The action of $\hat{\cal{S}}_x(-{\bm k})$ on the two-mode squeezed
state~(\ref{eq:qstate}) can be expressed as
\begin{align}
\hat{\cal{S}}_x(-{\bm k})&\vert \Psi_{2, {\mathrm{sq}}}\rangle 
=\int _0^{+\infty}{\rm d}\tilde{q}_{-{\bm k}}
\sum_{n=0}^{+\infty}
\frac{1}{\cosh r_k}e^{-2in\varphi_k}\tanh ^n r_k
\nonumber \\ & \times
\left(\langle \tilde{q}_{-{\bm k}}\vert n_{-{\bm k}}\rangle 
\vert n_{\bm k},\tilde{q}_{-{\bm k}}\rangle 
-
\langle -\tilde{q}_{-{\bm k}}\vert n_{-{\bm k}}\rangle 
\vert n_{\bm k},-\tilde{q}_{-{\bm k}}\rangle \right).
\end{align}
Then, the next step is to apply the operator $\hat{\cal{S}}_x({\bm k})$ on the 
previous state. This leads to 
\begin{widetext}
\begin{align}
\hat{\cal{S}}_x({\bm k})\hat{\cal{S}}_x(-{\bm k})\vert \Psi_{2, {\rm sq}}\rangle 
& =\int _0^{+\infty}\int _0^{+\infty}{\rm d}\tilde{q}_{-{\bm k}}
{\rm d}q_{{\bm k}}
\sum_{n=0}^{+\infty}
\frac{1}{\cosh r_k}e^{-2in\varphi_k}\tanh ^n r_k
\biggl(\langle \tilde{q}_{-{\bm k}}\vert n_{-{\bm k}}\rangle 
\langle q_{{\bm k}}\vert n_{{\bm k}}\rangle 
\vert q_{\bm k},\tilde{q}_{-{\bm k}}\rangle 
\nonumber \\ &
-\langle -\tilde{q}_{-{\bm k}}\vert n_{-{\bm k}}\rangle 
\langle q_{{\bm k}}\vert n_{{\bm k}}\rangle 
\vert q_{\bm k},-\tilde{q}_{-{\bm k}}\rangle 
-\langle \tilde{q}_{-{\bm k}}\vert n_{-{\bm k}}\rangle 
\langle -q_{{\bm k}}\vert n_{{\bm k}}\rangle 
\vert -q_{\bm k},\tilde{q}_{-{\bm k}}\rangle 
\nonumber \\ &
+\langle -\tilde{q}_{-{\bm k}}\vert n_{-{\bm k}}\rangle 
\langle -q_{{\bm k}}\vert n_{{\bm k}}\rangle 
\vert -q_{\bm k},-\tilde{q}_{-{\bm k}}\rangle 
\biggr),
\end{align}
and, therefore, using again the expression of the two-mode squeezed
state, one arrives at
\begin{align}
\langle \Psi_{2,{\rm sq}}\vert 
\hat{\cal{S}}_x({\bm k})\hat{\cal{S}}_x(-{\bm k})\vert \Psi_{2, {\rm sq}}\rangle 
& =\frac{1}{\cosh ^2r_k}
\sum_{n=0}^{+\infty}\sum_{m=0}^{+\infty}
e^{-2in\varphi_k+2im \varphi_k}\tanh ^n r_k\tanh ^m r_k
\int _0^{+\infty}\int _0^{+\infty}{\rm d}\tilde{q}_{-{\bm k}}
{\rm d}q_{{\bm k}}
\nonumber \\ & \times 
\biggl(\langle \tilde{q}_{-{\bm k}}\vert n_{-{\bm k}}\rangle 
\langle q_{{\bm k}}\vert n_{{\bm k}}\rangle 
\langle m_{\bm k}\vert q_{{\bm k}}\rangle 
\langle m_{-{\bm k}}\vert \tilde{q}_{-{\bm k}}\rangle 
-\langle -\tilde{q}_{-{\bm k}}\vert n_{-{\bm k}}\rangle 
\langle q_{{\bm k}}\vert n_{{\bm k}}\rangle 
\langle m_{\bm k}\vert q_{{\bm k}}\rangle 
\langle m_{-{\bm k}}\vert -\tilde{q}_{-{\bm k}}\rangle 
\nonumber \\ & 
-\langle \tilde{q}_{-{\bm k}}\vert n_{-{\bm k}}\rangle 
\langle -q_{{\bm k}}\vert n_{{\bm k}}\rangle 
\langle m_{\bm k}\vert -q_{{\bm k}}\rangle 
\langle m_{-{\bm k}}\vert \tilde{q}_{-{\bm k}}\rangle 
+\langle -\tilde{q}_{-{\bm k}}\vert n_{-{\bm k}}\rangle 
\langle -q_{{\bm k}}\vert n_{{\bm k}}\rangle 
\langle m_{\bm k}\vert -q_{{\bm k}}\rangle 
\langle m_{-{\bm k}}\vert -\tilde{q}_{-{\bm k}}\rangle
\biggr).
\end{align}
One can then express explicitly each of the matrix element present in
the above equation, obtaining the following expression
\begin{align}
\langle \Psi_{2,{\rm sq}}\vert 
\hat{\cal{S}}_x({\bm k})\hat{\cal{S}}_x(-{\bm k})\vert \Psi_{2, {\rm sq}}\rangle 
& =\frac{1}{\cosh ^2r_k}
\sum_{n=0}^{+\infty}\sum_{m=0}^{+\infty}
e^{-2in\varphi_k+2im \varphi_k}\tanh ^n r_k\tanh ^m r_k
\frac{1}{\sqrt{\pi}2^nn!}
\frac{1}{\sqrt{\pi}2^mm!}
\int _0^{+\infty}\int _0^{+\infty}{\rm d}\tilde{q}_{-{\bm k}}
{\rm d}q_{{\bm k}}
\nonumber \\ & 
\biggl[H_n(\tilde{q}_{-{\bm k}})
H_n(q_{{\bm k}})
H_m(q_{{\bm k}})
H_m(\tilde{q}_{-{\bm k}})
-H_n(-\tilde{q}_{-{\bm k}})
H_n(q_{{\bm k}})
H_m(q_{{\bm k}})
H_m(-\tilde{q}_{-{\bm k}})
\nonumber \\ & 
-H_n(\tilde{q}_{-{\bm k}})
H_n(-q_{{\bm k}})
H_m(-q_{{\bm k}})
H_m(\tilde{q}_{-{\bm k}})
+H_n(-\tilde{q}_{-{\bm k}})
H_n(-q_{{\bm k}})
H_m(-q_{{\bm k}})
H_m(-\tilde{q}_{-{\bm k}})
\biggr]e^{-\tilde{q}^2_{\bm{k}}-q^2_{\bm{k}}},
\end{align}
where we recall that $H_n(.)$ is a Hermite polynomial of order
$n$. This expression is made of four terms that can be calculated
separately. Using the relation
\begin{align}
\sum_{n=0}^{+\infty}\frac{w^n}{n!}H_n(x)H_n(y)
=\frac{1}{\sqrt{1-4w^2}}\exp\left\{\frac{2w}{4w^2-1}
\left[2w\left(x^2+y^2\right)-2xy\right]\right\},
\end{align}
the first term reads
\begin{align}
\langle \Psi_{2,{\rm sq}}\vert 
\hat{\cal{S}}_x({\bm k})\hat{\cal{S}}_x(-{\bm k})\vert \Psi_{2, {\rm sq}}\rangle ^{(1)}
=& \frac{1}{\pi \cosh ^2r_k}
\frac{1}{\sqrt{1-4w^2}}
\frac{1}{\sqrt{1-4w^*{}^2}}
\int _0^{+\infty}\int _0^{+\infty}{\rm d}\tilde{q}_{-{\bm k}}
{\rm d}q_{{\bm k}}
\nonumber \\ & \times
\exp\left\{-q_{\bm k}^2+\frac{2w}{4w^2-1}
\left[2w\left(q_{\bm k}^2+\tilde{q}_{-{\bm k}}^2\right)-2q_{\bm k}\tilde{q}_{-{\bm k}}
\right]\right\}
\nonumber \\ & \times
\exp\left\{-\tilde{q}_{\bm k}^2+\frac{2w^*}{4w^*{}^2-1}
\left[2w^*\left(q_{\bm k}^2+\tilde{q}_{-{\bm k}}^2\right)
-2q_{\bm k}\tilde{q}_{-{\bm k}}
\right]\right\},
\end{align}
where we have defined $w\equiv e^{-2i\varphi_k}\tanh r_k/2$. We see
that the only thing which remains to be calculated is a
two-dimensional Gaussian integral. Simplifying the argument of the
exponential, it takes the following form
\begin{align}
\label{eq:term1}
\langle \Psi_{2,{\rm sq}}\vert 
\hat{\cal{S}}_x({\bm k})\hat{\cal{S}}_x(-{\bm k})\vert \Psi_{2, {\rm sq}}\rangle ^{(1)}
=& \frac{1}{\pi \cosh ^2r_k}
\frac{1}{\sqrt{1-4w^2}}
\frac{1}{\sqrt{1-4w^*{}^2}}
\int _0^{+\infty}\int _0^{+\infty}{\rm d}\tilde{q}_{-{\bm k}}
{\rm d}q_{{\bm k}}\, 
e^{- A \left(\tilde{q}_{-{\bm k}}^2+q_{\bm k}^2\right)
+ C \tilde{q}_{-{\bm k}} q_{\bm k}}
\end{align}
where the coefficients $A$ and $C$ can be expressed as
\begin{align}
A & \equiv \frac{\tanh^2(r_k)+1}{\cosh^2(r_k)\left[\tanh^4(r_k)
-2\tanh^2(r_k)\cos(4\varphi_k)+1\right]}, \\
C &\equiv \frac{4\tanh(r_k)\cos(2\varphi_k)}{\cosh^2(r_k)\left[\tanh^4(r_k)
-2\tanh^2(r_k)\cos(4\varphi_k)+1\right]}.
\end{align}
The three other terms can also be calculated explicitly by using the
remark that $H_n(-x)=(-1)^nH_n(x)$. Then, one immediately sees that
the fourth term is in fact equal to the first one. It is also clear
that the second and the third ones are equal. The calculation of these
two last terms is very similar to the above calculation except that
the quantity $w$ should be substituted by $-w$ which leads to an
expression similar to \Eq{eq:term1}, the only difference being
that the sign of the term
$C \tilde{q}_{-{\bm k}} q_{\bm k}$ in the argument of the
exponential is changed. The final result reads
\begin{align}
\langle \Psi_{2,{\rm sq}}\vert 
\hat{\cal{S}}_x({\bm k})\hat{\cal{S}}_x(-{\bm k})\vert \Psi_{2, {\rm sq}}\rangle 
=& \frac{2}{\pi \cosh ^2r_k}
\frac{1}{\sqrt{1-4w^2}}
\frac{1}{\sqrt{1-4w^*{}^2}}
\nonumber \\ & \times
\int _0^{+\infty}\int _0^{+\infty}{\rm d}\tilde{q}_{-{\bm k}}
{\rm d}q_{{\bm k}}\, \left[
e^{- A \left(\tilde{q}_{-{\bm k}}^2+q_{\bm k}^2\right)
+ C \tilde{q}_{-{\bm k}} q_{\bm k}}
-e^{- A \left(\tilde{q}_{-{\bm k}}^2+q_{\bm k}^2\right)
- C \tilde{q}_{-{\bm k}} q_{\bm k}}\right].
\end{align}
The integration can be performed explicitly and one obtains
\begin{align}
\langle \Psi_{2,{\rm sq}}\vert 
\hat{\cal{S}}_x({\bm k})\hat{\cal{S}}_x(-{\bm k})\vert \Psi_{2, {\rm sq}}\rangle 
=& \frac{4}{\pi \cosh ^2r_k}
\frac{1}{\sqrt{1-4w^2}}
\frac{1}{\sqrt{1-4w^*{}^2}}
\frac{1}{\sqrt{4A^2-C^2}}\arctan\left(\frac{C}{\sqrt{4A^2-C^2}}\right)\, .
\end{align}
Finally, using the expression of $w$, one arrives at our final
expression, namely
\begin{align}
\label{eq:correlsxappendix}
\langle \Psi_{2,{\rm sq}}\vert 
\hat{\cal{S}}_x({\bm k})\hat{\cal{S}}_x(-{\bm k})\vert \Psi_{2, {\rm sq}}\rangle 
=& \frac{2}{\pi}\arctan\left[\frac{2\tanh(r_k)\cos(2\varphi_k)}
{\sqrt{\tanh^4(r_k)
-2\tanh^2(r_k)\cos(4\varphi_k)+1}}\right].
\end{align}
\end{widetext}
This is the expression used in the main text. Let us also notice that,
if the squeezing angle vanishes, $\varphi_k=0$, then one simply has
\begin{align}
\langle \Psi_{2,{\rm sq}}\vert 
\hat{\cal{S}}_x({\bm k})\hat{\cal{S}}_x(-{\bm k})\vert \Psi_{2, {\rm sq}}\rangle 
=& \frac{2}{\pi}\arctan\left[\sinh(2r_k)\right].
\end{align}
This expression is consistent with Eq.~(35) of
\Ref{2006FoPh...36..546R}, except that, in that reference, there is probably a
misprint in the expression of the Bell operator mean value which
should rather look like Eq.~(24) of the same article.

\bibliography{bell}

\end{document}